\pdfoutput=1

\documentclass[final,5p,times,twocolumn]{elsarticle}

\usepackage{graphicx,amssymb,amsmath,array,multirow,bm,subcaption}
\usepackage[nomarkers,nofiglist, notablist]{endfloat}

\graphicspath{{figures/}}

\begin{document}

\author[UW]{Tanyakarn Treeratanaphitak}
\author[UW]{Mark D Pritzker}
\ead{pritzker@uwaterloo.ca}
\author[UW,WIN]{Nasser Mohieddin Abukhdeir\corref{cor}}
\cortext[cor]{Corresponding author}
\ead[url]{http://chemeng.uwaterloo.ca/abukhdeir/}
\ead{nmabukhdeir@uwaterloo.ca}

\address[UW]{Department of Chemical Engineering}
\address[WIN]{Waterloo Institute for Nanotechnology\\University of Waterloo\\200 University Avenue West\\Waterloo, Ontario, Canada N2L 3G1}

\title{Kinetic Monte Carlo Simulation of Electrodeposition using the Embedded-Atom Method}

\biboptions{sort&compress}

\journal{Electrochimica Acta}

\begin{frontmatter}
\begin{abstract}
  A kinetic Monte Carlo (KMC) method for deposition is presented and applied to the simulation of electrodeposition of a metal on a single crystal surface of the same metal under galvanostatic conditions.
  This method utilizes the multi-body embedded-atom method (EAM) potential to characterize the interactions of metal atoms and adatoms.
  The method accounts for collective surface diffusion processes, in addition to nearest-neighbor hopping, including atom exchange and step-edge atom exchange.
  Steady-state deposition configurations obtained using the KMC method are validated by comparison with the structures obtained through the use of molecular dynamics (MD) simulations to relax KMC constraints.
  The results of this work support the use of the proposed KMC method to simulate electrodeposition processes at length (microns) and time (seconds) scales that are not feasible using other methods.
\end{abstract}


\end{frontmatter}

\section{Introduction}\label{sec:intro}
The effects of microstructure of metal films on device performance and longevity have become increasingly important as recent advances in the reduction of interconnect dimensions progress.
A specific example is the microstructure resulting from the copper damascene electroplating process \cite{Andricacos1998}.
Certain microscopic structures and interfaces between crystal grains in copper films have been found to improve the performance of interconnects \cite{Durkan2000,Hau-Riege2001}.
For example, `bamboo' grain structures and (111) orientation are preferred since they improve the lifetime of copper interconnects \cite{Hau-Riege2001}.
Thus, it is important to determine under which conditions the electrodeposition process yields these preferred structures.

A very effective method for simulation of the electrodeposition process, without resorting to first-principles calculations, is through the use of molecular dynamics (MD) that makes use of a suitable interaction potential.
The embedded-atom method (EAM) potential has been shown to accurately characterize metal/metal interactions \cite{Daw1984} and the predict relevant dynamics for systems including hydrogen adsorption onto nickel and segregation in binary alloys \cite{Daw1993}.
The EAM potential has been extensively validated for metallic systems \cite{Adams1989,Daw1984,Daw1993,Foiles1987} and used in MD simulations of hydrogen dissociation on nickel \cite{Foiles1987}, self-diffusion of metals \cite{Adams1989, Daw1993, Antczak2010} and epitaxial growth \cite{Mariscal2007}.
A significant limitation of MD is its computational requirement since it explicitly accounts for thermal fluctuations.
Thus, even with the use of parallel large-scale MD codes and a large number of parallel processors running over several days, simulations can only resolve time scales on the order of nanoseconds.
Even accelerated MD methods such as hyperdynamics \cite{Voter1997, Voter1997-2} and temperature-accelerated dynamics \cite{Sorensen2000-TAD, Voter2002} are limited to small systems.
Thus, an alternative method must be used to simulate phenomena over timescales on the order of seconds that are relevant to the electrodeposition process.
The method used in this work is kinetic Monte Carlo (KMC) \cite{Fichthorn1991}, which enables simulation over longer timescales with a much lower computational requirement.

Most of the previous research utilizing KMC to model the electrodeposition process has made use of the solid-on-solid (SOS) model developed by Gilmer and Bennema \cite{Gilmer1972}.
In the SOS model, the rates are a function of the number of occupied nearest neighbors with no vacancies allowed.
The interaction energy of the atoms and the system are not computed since no interaction potential is considered.
Hybrid multiscale simulation methods that blend the SOS technique (KMC) and continuum mechanics have been applied to model copper electrodeposition in trenches \cite{Zheng2008, Rusli2007}.
The SOS method has been applied to polycrystalline growth \cite{Rubio2003, Stephens2007}, facet growth \cite{Wang2000} and two-dimensional growth \cite{Liu2009, Liu2009EC, Liu2013}.  
However, a main drawback of the SOS model is that it does not accurately describe the metal crystalline microstructure, as is possible via MD using EAM potentials.

The SOS method also does not account for vacancies, motivating Kaneko and co-workers \cite{Kaneko2006, Kaneko2009, Kaneko2013} to introduce the solid-by-solid (SBS) method to address this limitation.
However, the SBS method suffers from the same limitations as the SOS in that it does not accurately describe the microstructure of the metal crystal.
Thus, structures obtained from SOS and SBS simulations are not always clearly related to a specific metal electrodeposition process.

In the present work, a KMC method (KMC-EAM) based on the highly descriptive EAM potential is presented which includes collective diffusion mechanisms (atom exchange and step-edge atom exchange), in addition to nearest-neighbor hopping.
In past work, MD simulations using the EAM potential (MD-EAM) have been used only to predict activation energies for KMC simulations using the SOS and SBS models \cite{Gilmer2000}.
Unlike the approach used in the SOS and SBS methods, the EAM potential is to form the Hamiltonian of the system in KMC-EAM simulations, not only to evaluate activation energies.
More recent past approaches using the EAM potential directly in Monte Carlo simulations have shown promising results for the simulation of electrodeposition \cite{Gimenez2002, Gimenez2003, Oviedo2005, Huang2009} including grand-canonical Monte Carlo and KMC simulations.
Gimenez et al. \cite{Gimenez2002} carried out KMC calculations using EAM potentials for two-dimensional deposition and limited the process to the growth of a single monolayer. 
Thus, these previous simulations were effectively limited to two-dimensional and sub-monolayer deposition dynamics.

In this work, the KMC-EAM method is applied to three-dimensional electrodeposition of a copper single crystal and validated by comparison with the equilibrium microstructures obtained by MD-EAM.
The MD-EAM method relaxes a number of the constraints and assumptions of the KMC-EAM method: the on-lattice approximation, finite diffusion mechanisms, and temporal coarse-graining.
The simulations are conducted over a range of current densities and temperatures that match common experimental conditions.
Simulations are then performed within these parameter ranges to predict the effect of current density and temperature on surface morphology.

The paper is organized as follows: Section \ref{sec:theory} -- background on KMC and the EAM potential, Section \ref{sec:methodology} -- presentation of the KMC-EAM method developed in this work, Section \ref{sec:results} -- simulation results on the effect of current and temperature on the accuracy of the KMC-EAM method and Section \ref{sec:concl} -- conclusions.

\section{Theory}\label{sec:theory}

\subsection{Kinetic Monte Carlo Method}\label{sec:KMC}

In MD, the exact locations of the atoms are determined and their motion is solved directly via Newton's equations of motion.
However, this is computationally expensive and so is limited to evolution of the domain over short time scales.
For metallic systems, it can be assumed that atoms vibrate about specific locations in quasi-equilibrium over a period of time.
Since each of these locations corresponds to a minimum in potential energy of the system, an atom must overcome an energy barrier to move from one minimum to another \cite{Chatterjee2007}.
Thus using a consistent fine-grained method, such as molecular dynamics or quantum mechanical density functional theory \cite{Chatterjee2007}, the ground state lattice type (FCC, BCC, etc) and lattice spacing of a specific atomic system \cite{Adams1989} are used as inputs for on-lattice KMC simulations.
This is the basis of the on-lattice approximation of for conducting KMC simulations of metal deposition via KMC \cite{Chatterjee2007}, whereby the metal atoms positions are limited only to sites on this crystal lattice.

Utilizing the on-lattice approximation, the discretized microscopic state $\bm{\sigma}$ of the system is a function of only lattice site occupancy and time where $\sigma_{i} = 0$ for a vacant site and $\sigma_{i} = 1$ for an occupied site.
In order to utilize the KMC methodology, an additional coarse-graining approximation must be used which assumes that the domain evolves through a discrete set of independent dynamic mechanisms.
Furthermore, these dynamic mechanisms are assumed to be Poisson processes \cite{Fichthorn1991}.
Given these approximations, the KMC method enables numerical solution of the Master equation of the system where the probability density $P(\bm{\sigma})$ of observing state $\bm{\sigma}$ is given as \cite{Fichthorn1991,Chatterjee2007}:
\begin{align}
\frac{\mathrm{d}P(\bm{\sigma})}{\mathrm{d}t} &= \sum_{\substack{\bm{\sigma}' \\ \bm{\sigma}' \neq \bm{\sigma}}}G(\bm{\sigma}' \rightarrow \bm{\sigma})P(\bm{\sigma}') - \sum_{\substack{\bm{\sigma'}\\ \bm{\sigma}' \neq \bm{\sigma}}}G(\bm{\sigma} \rightarrow \bm{\sigma}')P(\bm{\sigma}),
\label{eq:master_eqn1}
\end{align}
where $G(\bm{\sigma} \rightarrow \bm{\sigma'})$ is the probability per unit time that the system will undergo a transition from $\bm{\sigma}$ to $\bm{\sigma'}$.
Alternatively, \eqref{eq:master_eqn1} is also known as the \emph{chemical master equation} and may be reformulated as \cite{Gardiner1985}:
\begin{equation}
\mathrm{d}\sigma_{i} = \sum_{j}\Gamma_{ij}^{+}(\bm{\sigma})\mathrm{d}t - \sum_{j}\Gamma_{ij}^{-}(\bm{\sigma})\mathrm{d}t,
\label{eq:master_eqn2}
\end{equation}
where $\Gamma_{ij}(\bm{\sigma})$ is the transition probability ($\mathrm{s}^{-1}$) or propensity function for process $j$ at site $i$ when the state $\bm{\sigma}$ is observed.
The term $\Gamma_{ij}(\bm{\sigma})\mathrm{d}t$ gives the probability of state $\bm{\sigma}$ undergoing a change due to some move $j$ at site $i$ within the time increment $\mathrm{d} t$ \cite{Gillespie2007}.

\subsection{Embedded-Atom Method Potential}\label{sec:EAM}

The embedded-atom method potential is a semi-empirical potential that is based on density functional theory \cite{Daw1984}. 
This potential closely describes the effect of metallic bonding in metal systems to accurately estimate the potential energy of an atom \cite{Daw1993}.
The potential is composed of both multi-body and pairwise contributions \cite{Daw1984}:
\begin{equation}\label{eq:EAM}
E_{i} = F[\rho_{i}] + \frac{1}{2}\sum_{\substack{j \\ i \neq j}}\phi_{ij}(r_{ij})
\end{equation}
where $r_{ij}$ is the distance between atoms $i$ and $j$, $E_{i}$ is the interaction energy of atom $i$, $F$ is the multi-body \textit{embedding energy} functional and $\phi_{ij}(r_{ij})$ is a pair-wise repulsion between atoms $i$ and $j$.
The function $\rho_{i}$ is the total host electron density for atom $i$:
\begin{equation}\label{eq:EAM}
\rho_i =\sum_{\substack{j \\ i \neq j}}\rho_{h}(r_{ij})
\end{equation}
where $\rho_{h}$ is a function that quantifies the electron density of a neighboring atom.
The EAM parameters are estimated by fitting the EAM potential to known experimental values of metal properties such as the lattice constant, elastic constants, sublimation energy, and vacancy-formation energy \cite{Daw1984, Adams1989, Daw1993,Foiles1987}.

\section{Methodology}\label{sec:methodology}

The example chosen to apply and assess KMC-EAM in this work is copper electrodeposition onto a copper substrate (working electrode) from an acidic sulfate solution.
The overall reaction for the cathodic reduction of $Cu^{2+}$ is:
\begin{equation}
Cu_{(aq)}^{2+} + 2e^{-} \longrightarrow Cu_{(s)}^{0}.
\label{eq:cu_red}
\end{equation}
$Cu^{2+}$ ion reduction proceeds through consecutive single-electron transfer steps and involves the formation of an intermediate in which Cu has oxidation state $+1$ \cite{Mattsson1959, Conway1961}.
However, numerous studies have shown that the first of these steps has much slower kinetics than the second when copper deposition is carried out in acidic sulfate solutions \cite{Conway1961}.
Thus the first step is rate-determining \cite{Mattsson1959, Conway1961} and the two steps effectively occur almost simultaneously under these conditions.
In this study, the deposition mechanism is assumed to be kinetically controlled. Thus, transport of $Cu^{2+}$ within the solution to the electrode surface has no influence on the deposition rate and so only phenomena occurring on the copper surface are considered in the model and simulations.

The EAM interaction potential parameters for copper are taken from Adams et al \cite{Adams1989}.
This potential is expressed as a function of the atom separation distance in the form of cubic splines, one for the embedding term and one for the pair-wise repulsion term.
The energy of each atom is obtained by interpolating these splines according to the separation distance between the atom and each of its neighbors for both embedding energy and pair-wise repulsion contributions to the EAM potential.
The neighbor contribution is limited to atoms within a cutoff distance of 0.495 nm, as is consistent with EAM parameters obtained from Adams et al \cite{Adams1989}.

The lattice used for KMC-EAM simulations is consistent with the EAM parameters for copper.
This lattice type is FCC with a lattice spacing of 0.3615 nm which was determined experimentally  and was one of the properties that the EAM potential was fitted to \cite{Adams1989}.
Thus combining the on-lattice approximation and EAM potential, the Hamiltonian for the system is:
\begin{equation}\label{eq:hamilt}
E = \sum_{i=1}^{N} \sigma_{i} E_{i} = \sum_{i=1}^{N} \sigma_{i} \left(F[\rho_{i}] + \frac{1}{2}\sum_{j \neq i}^{n}\phi_{ij}(r_{ij}) \right)
\end{equation}
where $N$ is the number of sites in the system, $\sigma_{i}$ is the occupancy of site $i$ ($0$ or $1$), $j$ is a neighbor of site $i$ that is within the cutoff distance and $n$ is the number of sites within the cutoff distance.

\subsection{Processes}\label{sec:mech}

In this work, two dynamic processes are considered in modeling copper electrodeposition: (i) reduction of metal ions and deposition onto the surface as adsorbed adatoms and (ii) diffusion of these adatoms on the surface.
Diffusion in the bulk of the electrode is not considered since simulations are performed under conditions in which very few vacancies form \cite{Wolf1992}.
Lattice relaxation mechanisms are not considered because the on-lattice approximation is used.
Diffusion of adatoms on the deposit surface is complex and involves several collective mechanisms (concerted mechanisms) \cite{Antczak2010}, in addition to nearest-neighbor hopping.  

\begin{figure}
        \centering
        \begin{subfigure}[b]{0.25\linewidth}
                \centering
                \includegraphics[width=\linewidth]{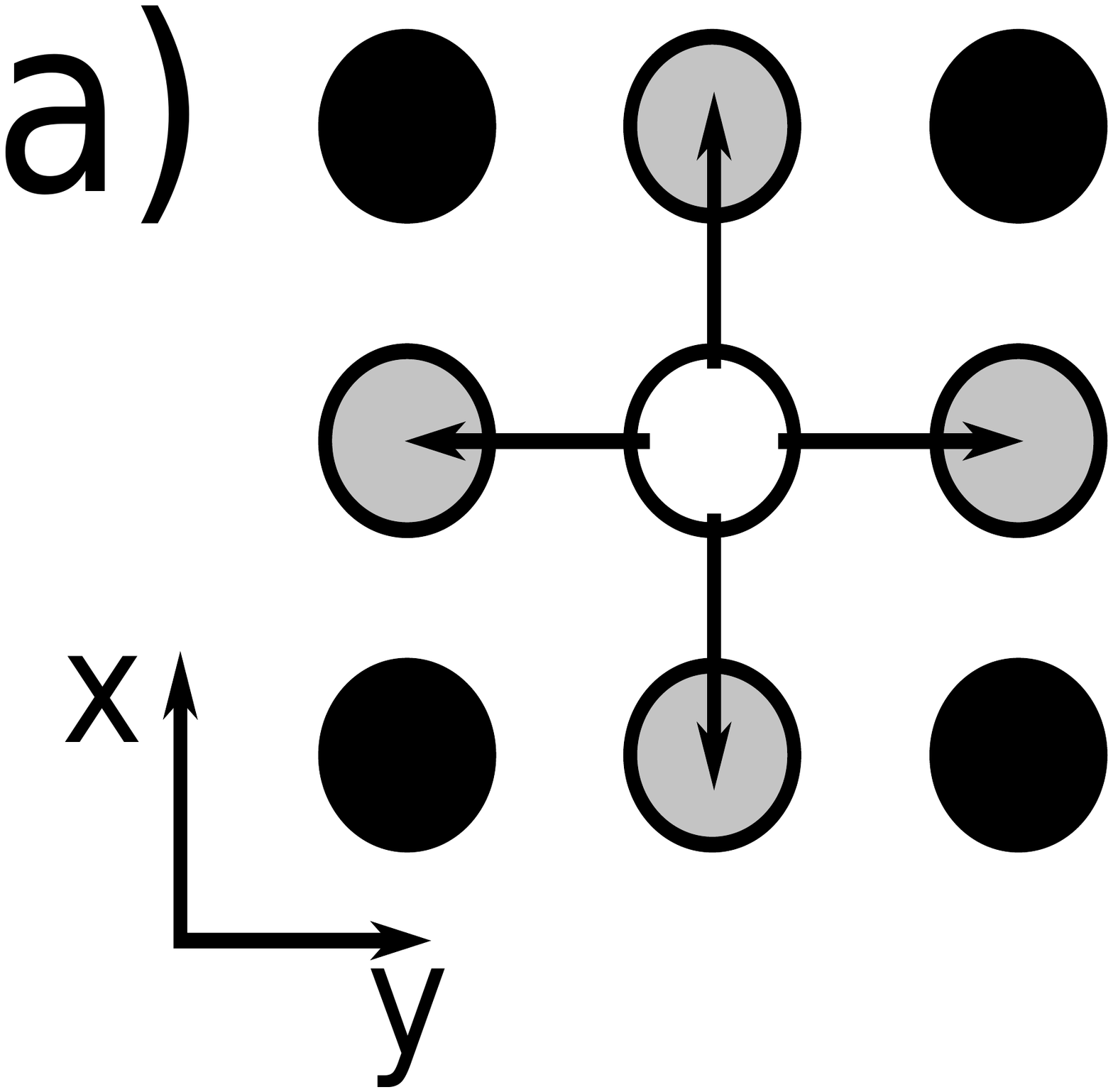}
                \caption{}
        \end{subfigure}%
        ~
        \begin{subfigure}[b]{0.33\linewidth}
                \centering
                \includegraphics[width=0.8\linewidth]{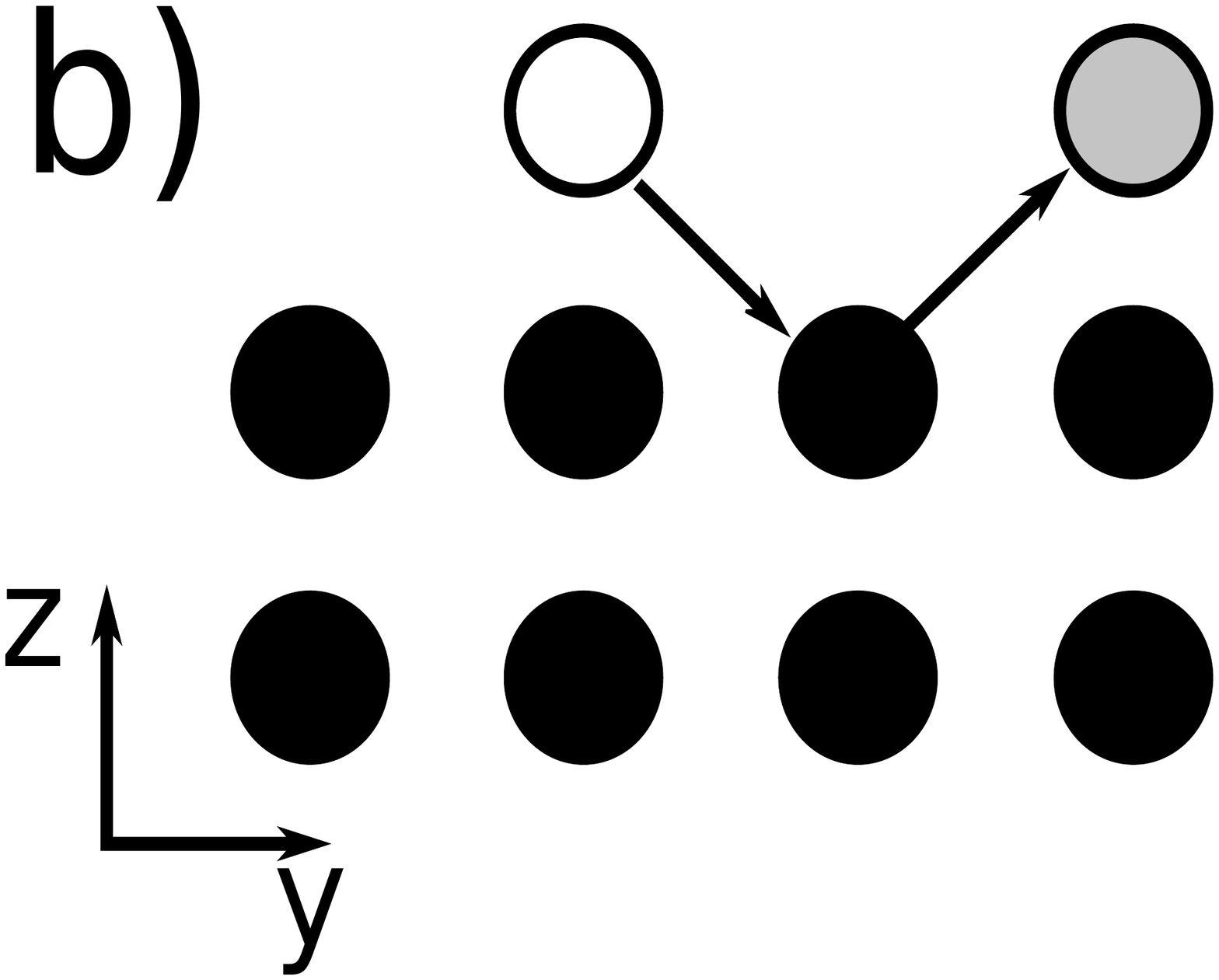}
                \caption{}
        \end{subfigure}
        ~
        \begin{subfigure}[b]{0.33\linewidth}
                \centering
                \includegraphics[width=0.8\linewidth]{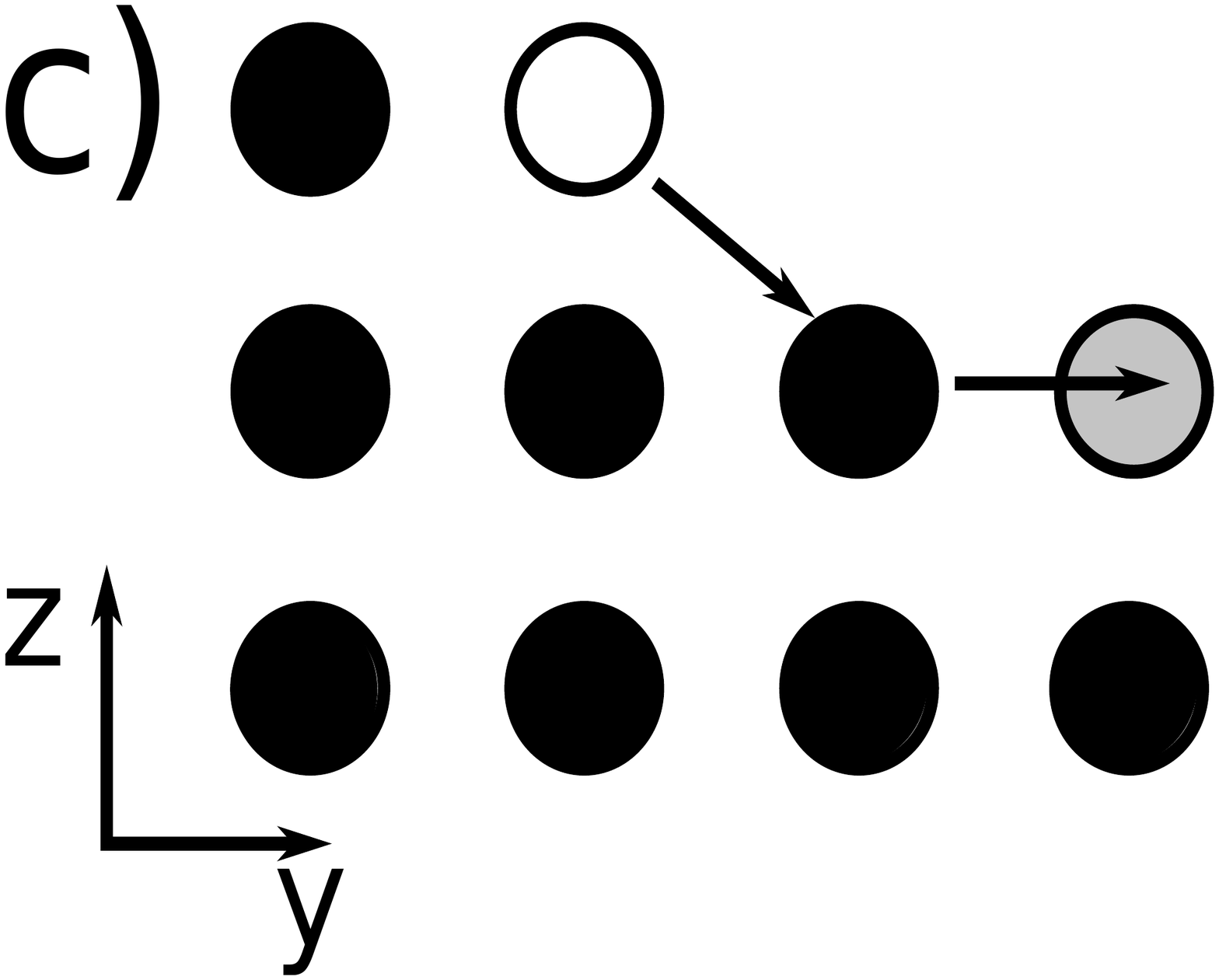}
                \caption{}
        \end{subfigure}
        \caption{Schematics of three possible diffusion mechanisms: a) hopping b) atom exchange and c) step-edge atom exchange. White denotes an adatom, grey denotes the new location of the adatom, and black denotes an occupied site. Note that the atom locations are not drawn to scale.\label{fig:diff_mech}}
\end{figure}

Three possible adatom surface diffusion mechanisms (shown in Figure \ref{fig:diff_mech}) are included in the model: hopping (single), atom exchange (collective) and step-edge atom exchange (collective).
Hopping (Figure \ref{fig:diff_mech}a) involves the diffusion of single adatoms and kink atoms, identified by coordination number $\le 6$ \cite{Budevski1996}, to unoccupied nearest-neighbor sites.
Most previous simulations include only this mechanism \cite{Rusli2007, Rubio2003, Liu2009, Liu2009EC, Liu2013, Battaile2008, Stephens2007}.

Atom exchange (Figure \ref{fig:diff_mech}b) involves the simultaneous (i) displacement of a sub-surface crystalline atom by a nearest-neighbor adatom and (ii) the hopping of the sub-surface crystalline atom to an unoccupied nearest-neighbor site at the surface (eq. \eqref{eq:atom_exch}).
Thus, the adatom and sub-surface crystalline atom exchange states -- the sub-surface atom becomes an adatom, while the adatom becomes part of the bulk \cite{Antczak2010}.

A special case of atom exchange occurs when the exchange occurs at the edge of a terrace/step in the surface; this atom exchange process is called step-edge atom exchange (Figure \ref{fig:diff_mech}c).
Unlike the previously described atom exchange mechanism, the sub-surface atom hops horizontally within the same layer.
The adatom becomes part of the surface crystal and sub-surface atom becomes either an adatom or a kink site depending on the coordination number of its new site.

The propensity functions for each type of diffusion event, used in the KMC-EAM method, are given in Table \ref{tab:trans_prob} \cite{Battaile2008}.
The numerical values of the parameters contained in the propensity functions used in this work are given in Table \ref{tab:parameters}.
The deposition propensity (eq. \eqref{eq:dep_rate}) is based on the relationship between the partial current density ($i_{dep}$) and deposition frequency given by Budevski et al \cite{Budevski1996}.
Furthermore, simulations are restricted to copper deposition occurring at low enough currents that transport of $Cu^{2+}$ from the electrolyte to the cathode has no influence on the process.

\begin{table*}
\caption{Propensity functions for the possible events}
\centering
\begin{tabular}{m{2.75cm} c}
\firsthline
Mechanism & Propensity Function \\ 
\hline
Deposition & \parbox{0.8\linewidth}{
\begin{equation}\Gamma_{i,dep} = \frac{i_{dep}}{-zen_{dep}} \label{eq:dep_rate}
\end{equation}} \\ 
\hline 
Hopping & \parbox{0.8\linewidth}{\begin{equation}
\Gamma_{i,hop} = \begin{cases}
                \nu \exp\left(-\frac{E_{hop}}{k_{B}T}\right) & \Delta E \leq 0 \\
                \nu \exp\left(-\frac{E_{hop}+\Delta E}{k_{B}T}\right) & \Delta E > 0
                \end{cases}
\label{eq:hopping}
\end{equation}} \\ 
\hline 
Atom exchange & \parbox{0.8\linewidth}{\begin{equation}
\Gamma_{i,exch} = \begin{cases}
                \nu \exp\left(-\frac{E_{exch}}{k_{B}T}\right) & \Delta E \leq 0 \\
                \nu \exp\left(-\frac{E_{exch}+\Delta E}{k_{B}T}\right) & \Delta E > 0
                \end{cases}
\label{eq:atom_exch}
\end{equation}} \\ 
\hline 
Step edge \newline atom exchange & \parbox{0.8\linewidth}{\begin{equation}
\Gamma_{i,step} = \begin{cases}
                \nu \exp\left(-\frac{E_{step}}{k_{B}T}\right) & \Delta E \leq 0 \\
                \nu \exp\left(-\frac{E_{step}+\Delta E}{k_{B}T}\right) & \Delta E > 0
                \end{cases}
\label{eq:step_edge}
\end{equation}} \\ 
\lasthline 
\end{tabular}
\label{tab:trans_prob}
\end{table*}

\begin{table*}
\caption{Parameters used in propensity functions for KMC-EAM.}
\centering
\begin{tabular}{l l l}
\firsthline
Parameter & Definition & Value \\
\hline
$n_{dep}$ & number of possible deposition sites per unit area & varies [=] sites $\mathrm{m}^{-2}$ \\
$e$ & elementary charge & $1.602 \times 10^{-19}$ C \\
$z$ & number of electrons transferred in reduction reaction & 2 \\
$\nu$ & atomic vibrational frequency& $2 \times 10^{13}$ $\mathrm{s}^{-1}$ \\
$E_{hop}$ & hopping activation energy & 0.5 eV \cite{Antczak2010} \\
$E_{exch}$ & atom exchange activation energy & 0.7 eV \cite{Antczak2010} \\
$E_{step}$ & step-edge atom exchange activation energy & 0.2 eV \cite{Antczak2010} \\
\lasthline
\end{tabular}
\label{tab:parameters}
\end{table*}

\subsection{Simulation Conditions}\label{sec:cond}

KMC-EAM simulations are carried out for a slab geometry that is infinite in the $x-y$ plane on which deposition occurs and semi-infinite in the $z$ direction normal to this plane.
Periodic boundary conditions are assumed in the $x-y$ plane to approximate an infinite plane. 
The copper substrate surface lies along the $\lbrace100\rbrace$ family of planes.
In addition to the process and material parameters presented above, input parameters for the simulations include the initial copper substrate seed layer height $h_{s}$ and the occupancy fraction $f_{s}$. 
The simulation domain sizes used range from $25a \times 25a \times 15a$ to $50a \times 50a \times 15a$ ($a_{Cu} = 0.3615~\mathrm{nm}$ is the lattice constant of copper).

During the first stage of the simulation, $2.5\times10^{4}$ atoms are deposited at different deposition rates and allowed to diffuse.
Following the deposition of all the atoms, simulation continues (in the absence of further deposition) until the system reaches equilibrium.
Equilibrium is identified when the change of the mean energy of the system with respect to time approaches zero with a tolerance of $1\%$.

The equilibrium configuration predicted by KMC-EAM in each case is evaluated by comparing it to the configuration obtained from a simulation using an established MD-EAM method.
This is done to validate the equilibrium state obtained from KMC-EAM and not the dynamics predicted by KMC-EAM.
This MD-EAM simulation uses the equilibrium configuration predicted by KMC-EAM as its initial condition and involves no further deposition to relax the constraints imposed by KMC-EAM as described in Section \ref{sec:intro}. 
The MD-EAM simulations are carried out using the canonical ensemble (constant number of atoms, volume and temperature) at the same temperature as the corresponding KMC-EAM simulation over a period of 6 nanoseconds.
The resulting configuration is then compared to that from KMC-EAM on the basis of the (i) equilibrium energy per atom and (ii) average coordination number.

The KMC simulation package that is the basis of the method is the Stochastic Parallel Particle Kinetic Simulator (SPPARKS:spparks.sandia.gov) \cite{Plimpton2009}.
The Gibson-Bruck \cite{Gibson2000} implementation of the direct Gillespie method is used to evolve the system.
The MD simulation package used for comparisons of equilibrium deposits is the Large-scale Atomic/Molecular Massively Parallel Simulator (LAMMPS:lammps.sandia.gov) described in Plimpton et al \cite{Plimpton1995}.

\section{Results and Discussion}\label{sec:results}

KMC-EAM simulations are performed to model deposition of a fixed number ($2.5\times10^{4}$) of copper atoms for different sets of initial conditions which govern deposition rates.
These initial conditions include domain size, thickness of the substrate layer and occupancy fraction of the substrate layer. 
Once deposited, the atoms are allowed to diffuse over the surface via the three mechanisms described in Table \ref{tab:trans_prob}. 

\begin{figure*}
        \centering
        \begin{subfigure}[b]{0.45\linewidth}
                \centering
                \includegraphics[width=\linewidth]{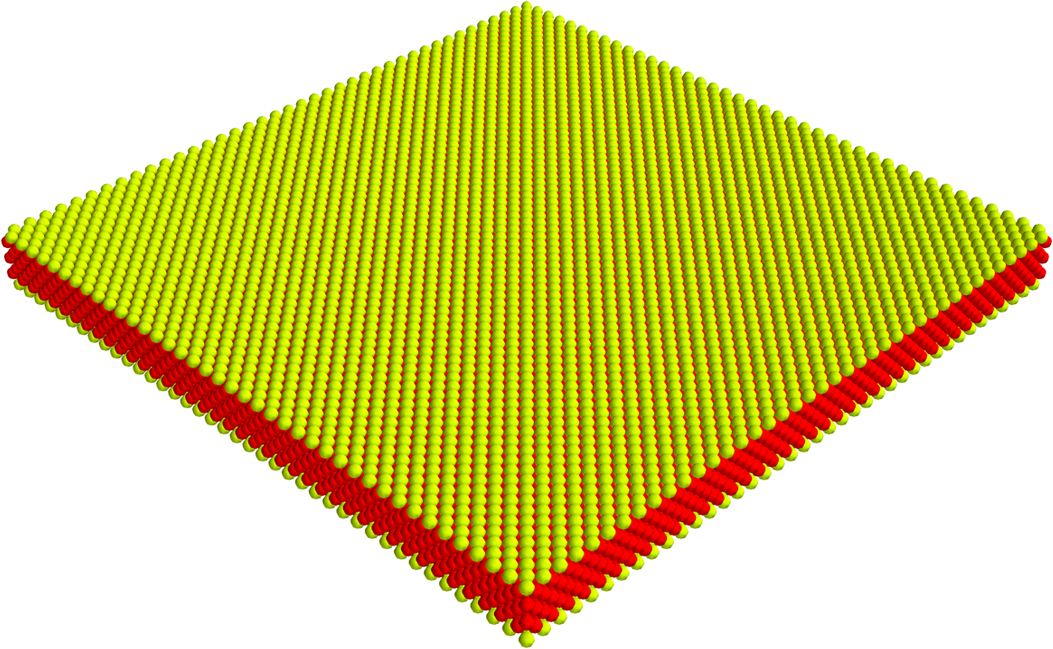}
                \caption{}
        \end{subfigure}%
        \begin{subfigure}[b]{0.45\linewidth}
                \centering
                \includegraphics[width=\textwidth]{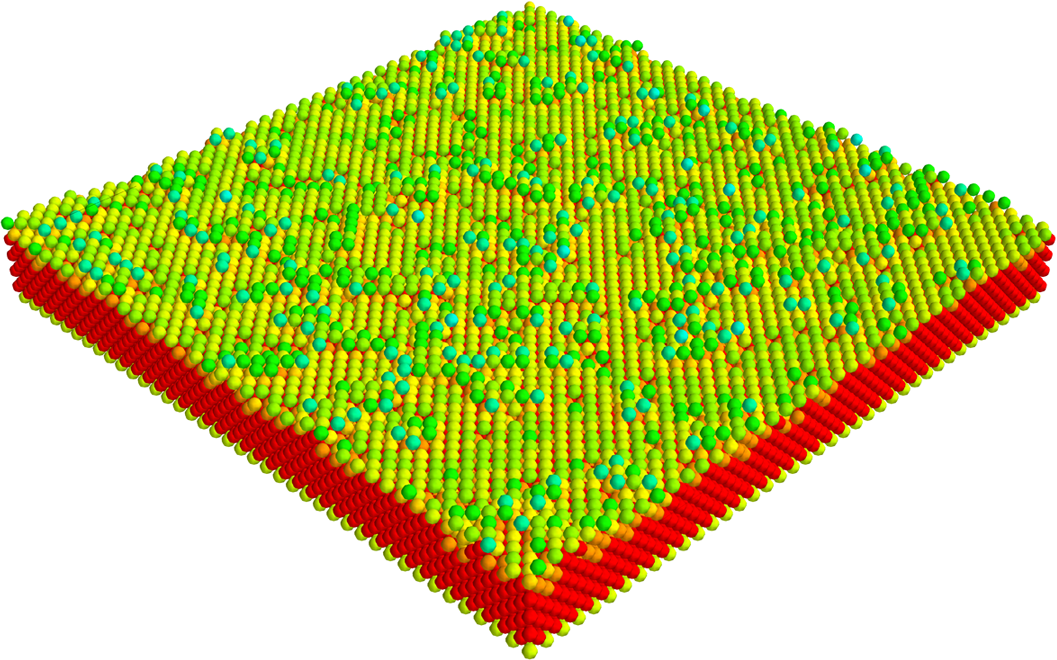}
                \caption{}
        \end{subfigure}
        
        \begin{subfigure}[b]{0.45\linewidth}
                \centering
                \includegraphics[width=\linewidth]{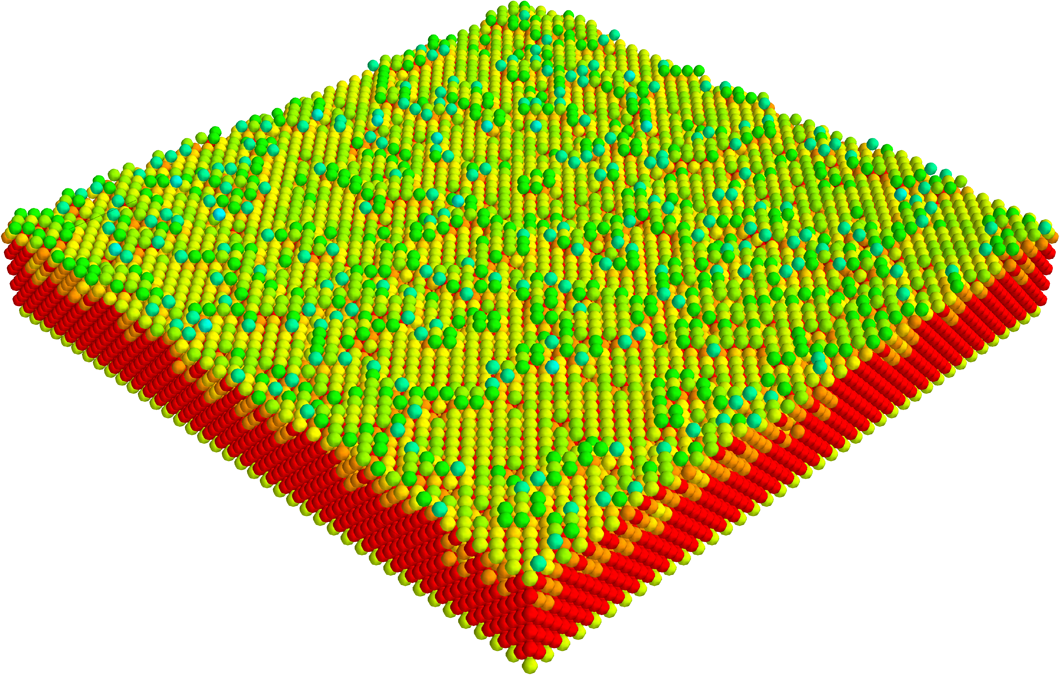}
                \caption{}
        \end{subfigure}
        \begin{subfigure}[b]{0.45\linewidth}
                \centering
                \includegraphics[width=\linewidth]{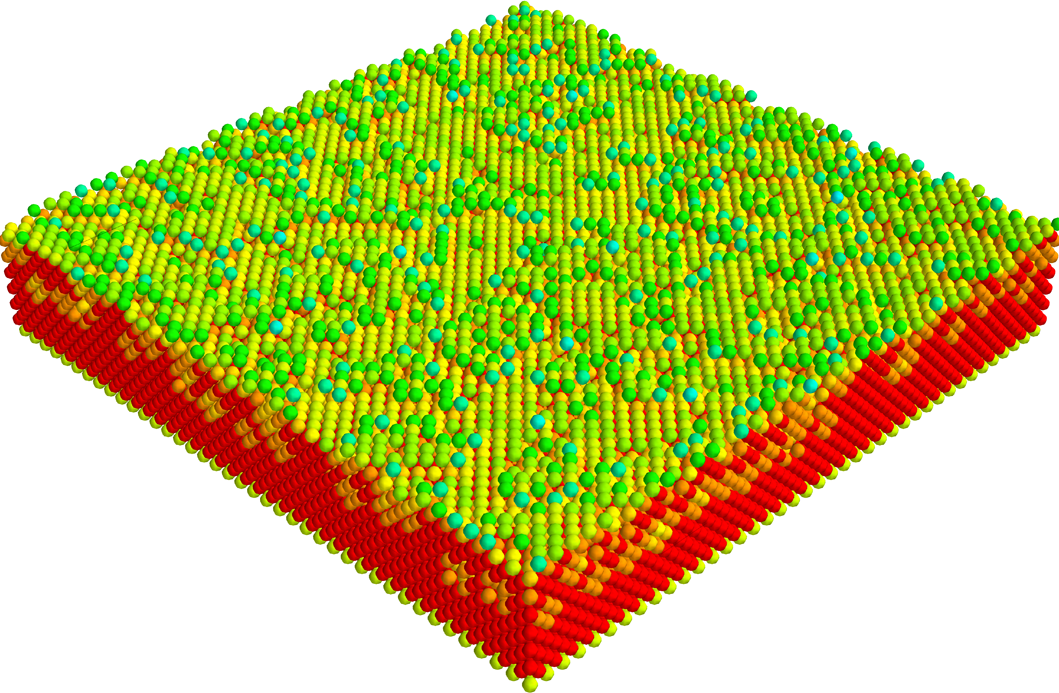}
                \caption{}
        \end{subfigure}

        \begin{subfigure}[b]{0.45\linewidth}
                \centering
                \includegraphics[width=\linewidth]{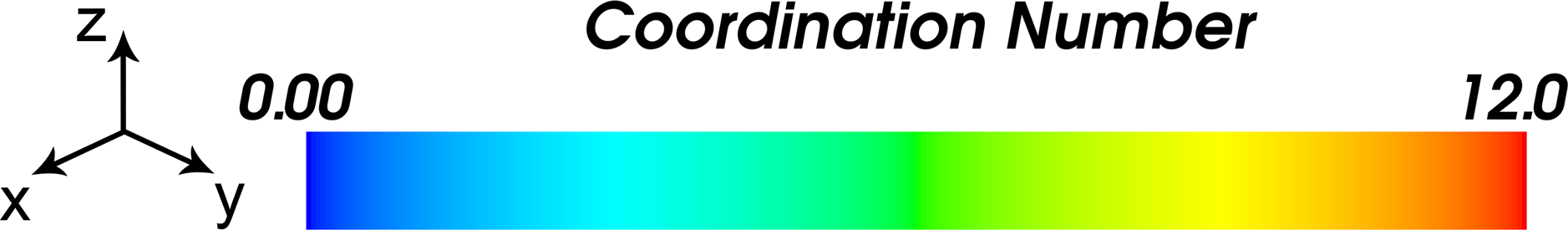}
        \end{subfigure}
        \caption{Morphology evolution of the configuration from KMC-EAM simulation at a) 0 s, b) 0.1 s, c) 0.2 s, and d) 0.3 s.  The current density is $-1000~\mathrm{A}~\mathrm{m}^{-2}$ and the operating temperature is 300 K.  Colors denote coordination number (blue to red in ascending order). The substrate surface area is $40a_{Cu} \times 40a_{Cu}$ ($\approx 210$ nm\textsuperscript{2}).\label{fig:snapshots}}
\end{figure*}

Simulations are conducted over a range of deposition current densities and operating temperatures. 
Temperatures between $300-330~\mathrm{K}$ are considered to span typical operating conditions used in industry and experimental studies. 
Current densities ranging from $-10~\mathrm{A}~\mathrm{m}^{-2}$ to $-1000~\mathrm{A}~\mathrm{m}^{-2}$ are chosen to span conditions from low to high deposition rates. 
This study is restricted to conditions where the deposition rate is kinetically controlled and unaffected by mass transfer. 
The $Cu^{2+}$ concentration in the bulk is assumed to be $1~\mathrm{mol}~\mathrm{dm^{-3}}$ to ensure that deposition remains in the kinetically-controlled regime for all current densities applied in the simulations.
At this bulk concentration, the highest current density of $-1000~\mathrm{A}~\mathrm{m}^{-2}$ considered is less than 20\% of the limiting current density for copper deposition onto a disk electrode rotating at 1000 RPM, as estimated using the Levich equation \cite{Vazquez2013,Bard2001}:
\begin{equation}
i_{L} = 0.620zFD^{2/3}w^{1/2}\nu^{-1/6}c.
\label{eq:levich}
\end{equation}
In this expression, $i_{L}$ is the limiting current density ($\mathrm{A}~\mathrm{m}^{-2}$), $D$ is the diffusion coefficient ($\mathrm{m^{2}}~\mathrm{s^{-1}}$), $w$ is the angular rotation speed ($\mathrm{rad}~\mathrm{s^{-1}}$), $\nu$ is the kinematic viscosity ($\mathrm{m^{2}}~\mathrm{s^{-1}}$) and $c$ is the concentration of the plating bath ($\mathrm{mol}~\mathrm{m^{-3}}$).
The initial occupancy fraction $f_{s}$ in the substrate layer is taken to be 1.0 in every simulation, while the initial copper substrate layer height $h_{s}$ is set to $1.1~\mathrm{nm}$.
Sample electrodeposition deposit morphology evolution from a KMC-EAM simulation is shown in Figure \ref{fig:snapshots}.

The first set of results focuses on the influence of the different surface diffusion mechanisms considered in the KMC-EAM method on the resulting deposit roughness and nanoscale morphology.
In particular, a comparison is made between the coatings obtained when surface diffusion occurs by hopping alone to those obtained when all three surface diffusion mechanisms operate.
Equilibrium deposit morphologies were characterized using mean roughness and local morphological measures -- area, perimeter and average curvature.
More detail on the evaluation of these morphological quantities and their meaning for deposit surfaces is provided in Section \ref{sec:effect_diff_mech}.

The second set of results involves the use of equilibrium deposit configurations from KMC-EAM, which correspond to electrodeposition over experimentally relevant timescales (seconds), as initial conditions for MD-EAM simulations.
These MD-EAM simulations runs were used to determine the approximation error associated with the assumptions required for KMC-EAM -- the on-lattice approximation, limitation of diffusion mechanisms, and time-coarse graining. 
Variation of deposition rate and temperature on the accuracy of the KMC-EAM was then determined in this way.

In order to characterize the kinetics of the deposition process, the mean energy and average coordination number of the configurations are used.
A consideration in comparing KMC-EAM and MD-EAM results is that KMC does not explicitly account for the average kinetic energy of the atoms.
Thus, the potential energy contribution to the total energy from MD-EAM is compared to the mean energy from KMC.
The average absolute relative energy difference per atom ($\delta_{E}$) and average absolute relative coordination number difference ($\delta_{C}$) between KMC-EAM and MD-EAM are used as a measure of how equilibrium configurations from the KMC-EAM method compare to equilibrium configurations from MD-EAM.
The average root-mean-squared displacement per atom (RMS displacement) in MD-EAM simulations is utilized as a means of tracking the distance atoms travel from their starting configuration, which corresponds to the equilibrium configuration from KMC-EAM.

\subsection{Kinetics of Diffusion Events}

\begin{figure}
\centering
        \begin{subfigure}[b]{0.5\linewidth}
                \centering
                \includegraphics[width=\linewidth]{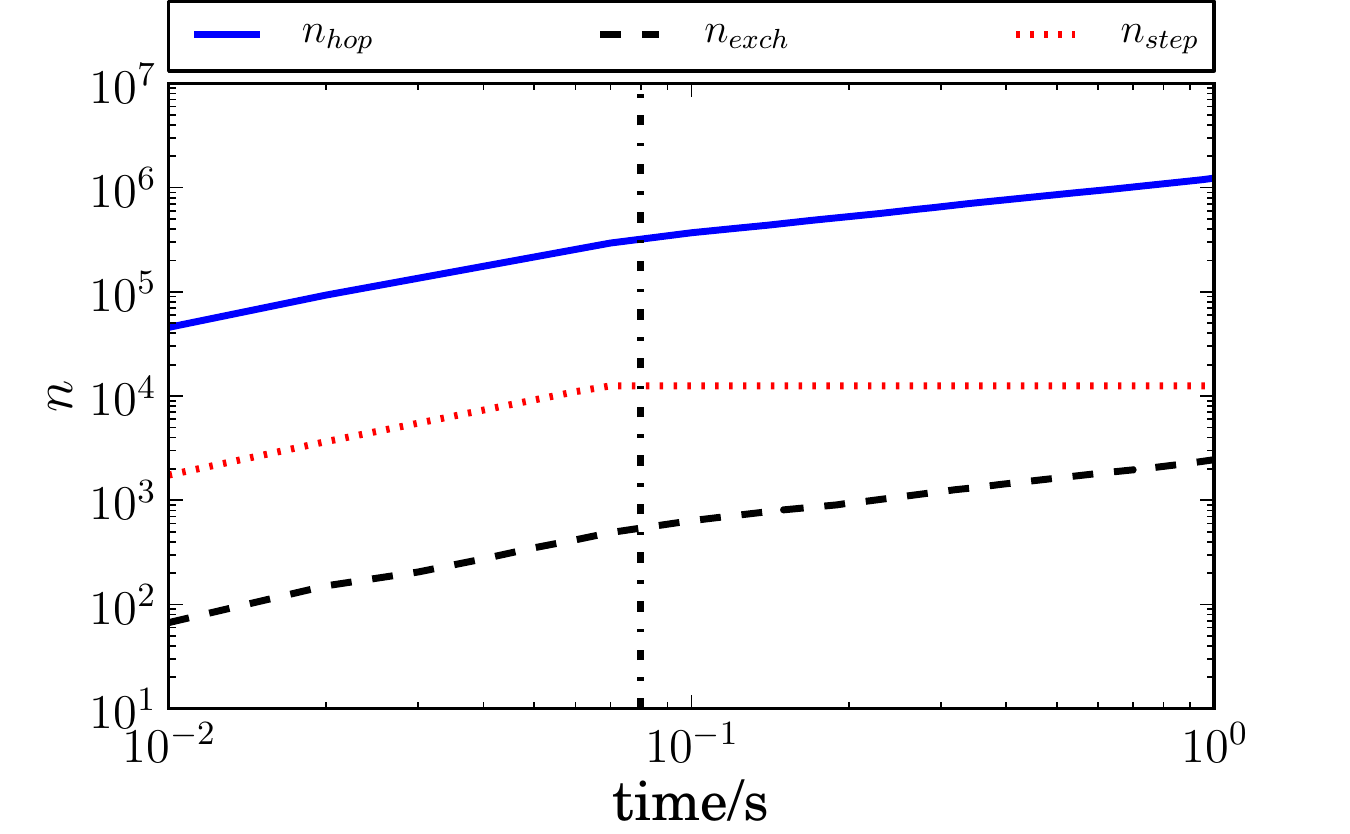}
                \caption{}
        \end{subfigure}%
        \begin{subfigure}[b]{0.5\linewidth}
                \centering
                \includegraphics[width=\linewidth]{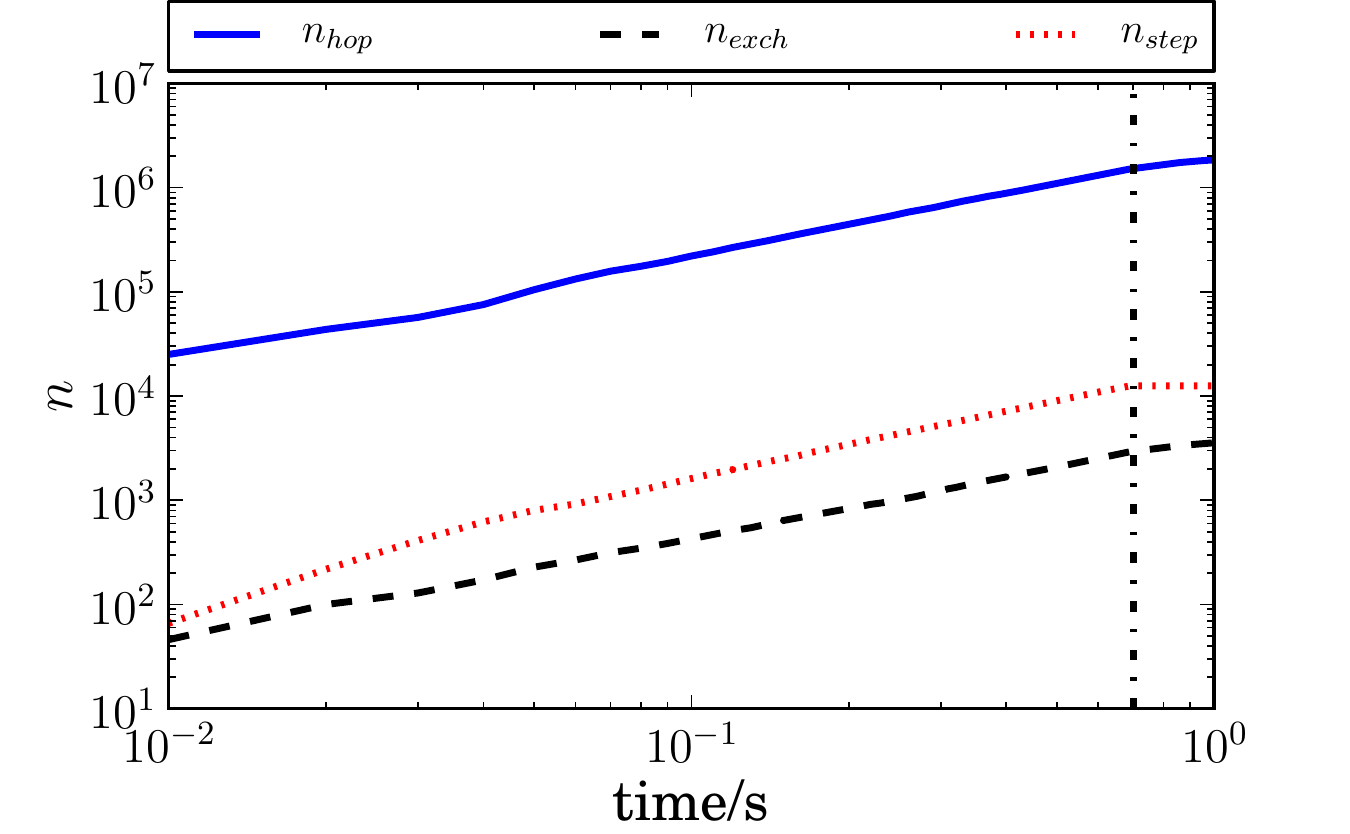}
                \caption{}
        \end{subfigure}%
    \caption{Number of diffusion events ($n$) over time at a) 300 K and $-1000~\mathrm{A}~\mathrm{m}^{-2}$ and b) 300 K and $-100~\mathrm{A}~\mathrm{m}^{-2}$.\label{fig:events}}
\end{figure}

Figure \ref{fig:events} shows the cumulative number of diffusion moves for each diffusion mechanism versus time for the first $1~\textrm{s}$ of simulation time for two different current densities ($-1000~\mathrm{A}~\mathrm{m}^{-2}$ and $-100~\mathrm{A}~\mathrm{m}^{-2}$).
In both simulations, all diffusion mechanisms are active during the electrodeposition phase.
Following the cessation of deposition (denoted with vertical line in Figure \ref{fig:events}), the step-edge atom exchange diffusion mechanism (Figure \ref{fig:diff_mech}c) ceases in both simulations while both the hopping (Figure \ref{fig:diff_mech}a) and atom exchange (Figure \ref{fig:diff_mech}b) surface diffusion mechanisms persist.
Both in the initial electrodeposition and equilibrium regime, diffusion events are observed to have a power law relationship with respect to time as indicated by the linear trends in Figures \ref{fig:events}a-b.
This implies that growth of the deposit surface occurs in a self-similar way where deposit morphology is consistent as film thickness increases.
Following deposition growth, the step-edge atom exchange mechanism ceases which indicates that only hopping and atom exchange diffusion mechanisms are important in the equilibrium regime.

The hopping surface diffusion mechanism is found to be dominant both in the growth and equilibrium regimes.
Any adatom can undergo hopping on the surface, while only atoms that satisfy the restrictions outlined in Section \ref{sec:mech} can undergo atom exchange and step-edge atom exchange.
Given that there are restrictions on the sites that atom exchange and step-edge atom exchange surface diffusion could occur at, the fact that hopping is the most frequent event is expected.

The step-edge atom exchange is found to be only present during the growth regime, which is reasonable given that the mechanism results in a new configuration that precludes the possibility of the event happening again in that locality with respect to the atoms undergoing the exchange.
Given the conditions for the mechanism (Figure \ref{fig:diff_mech}), diffusion via this mechanism ceases when deposition has stopped because no additional step-edges are being created. 
The duration of time during which the step-edge atom exchange mechanism is most active depends on the current density which determines the rate of deposition. 
As the deposition rate is increased the interval over which the step-edge exchange mechanism is most active decreases. 
This is supported by increase in slope of $n_{step}$ during the deposition stage at a current density of $-1000~\mathrm{A}~\mathrm{m}^{-2}$ versus that at $-100~\mathrm{A}~\mathrm{m}^{-2}$ (Figure \ref{fig:events}). 
Alternatively, the maximum value of $n_{step}$ is independent of deposition rate, comparing Figures \ref{fig:events}a and \ref{fig:events}b. 
Instead, the value of $n_{step}$ at any time is related, primarily, to the total number of atoms deposited up to that point.

\begin{figure*}
        \centering
        \begin{subfigure}[b]{0.45\linewidth}
                \centering
                \includegraphics[width=\linewidth]{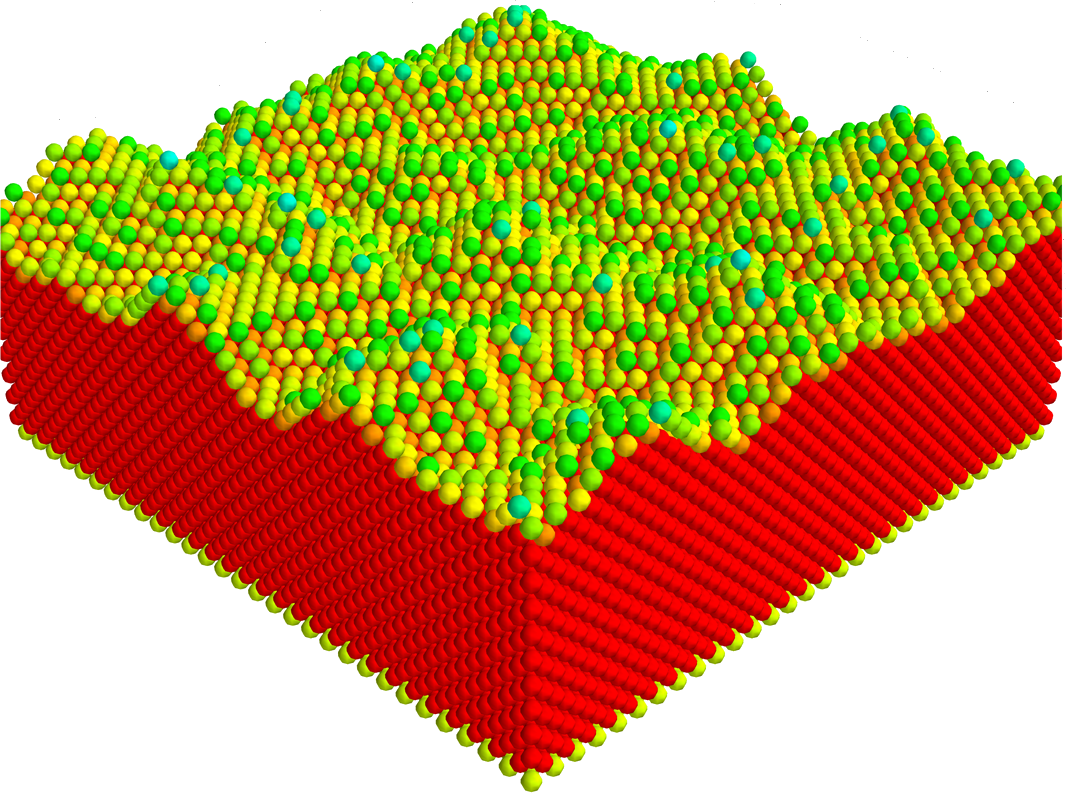}
                \caption{}
        \end{subfigure}%
        \begin{subfigure}[b]{0.45\linewidth}
                \centering
                \includegraphics[width=\linewidth]{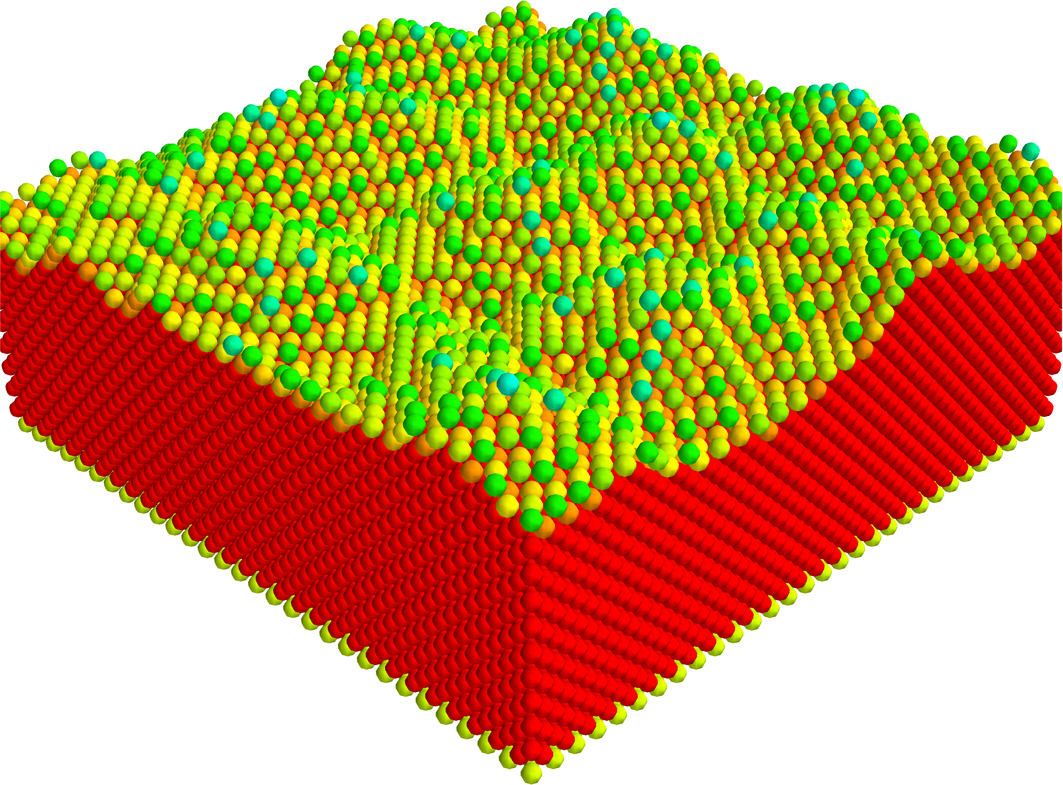}
                \caption{}
        \end{subfigure}
        
        \begin{subfigure}[b]{0.45\linewidth}
                \centering
                \includegraphics[width=\linewidth]{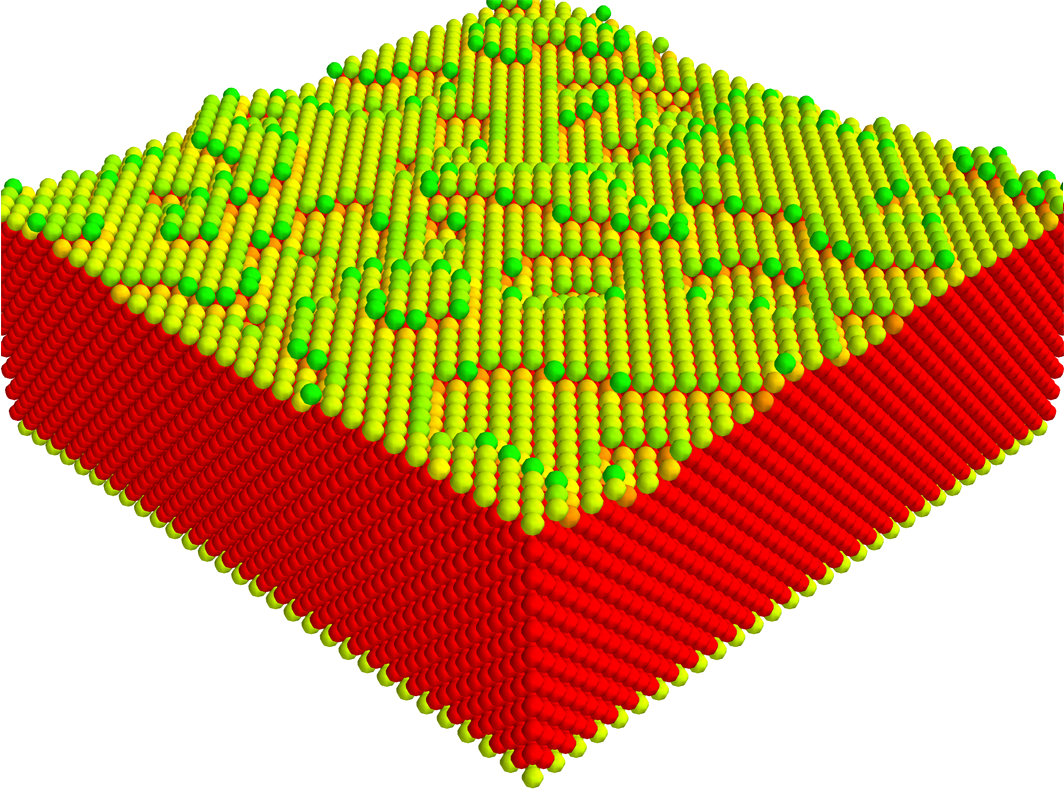}
                \caption{}
        \end{subfigure}
        \begin{subfigure}[b]{0.45\linewidth}
                \centering
                \includegraphics[width=\linewidth]{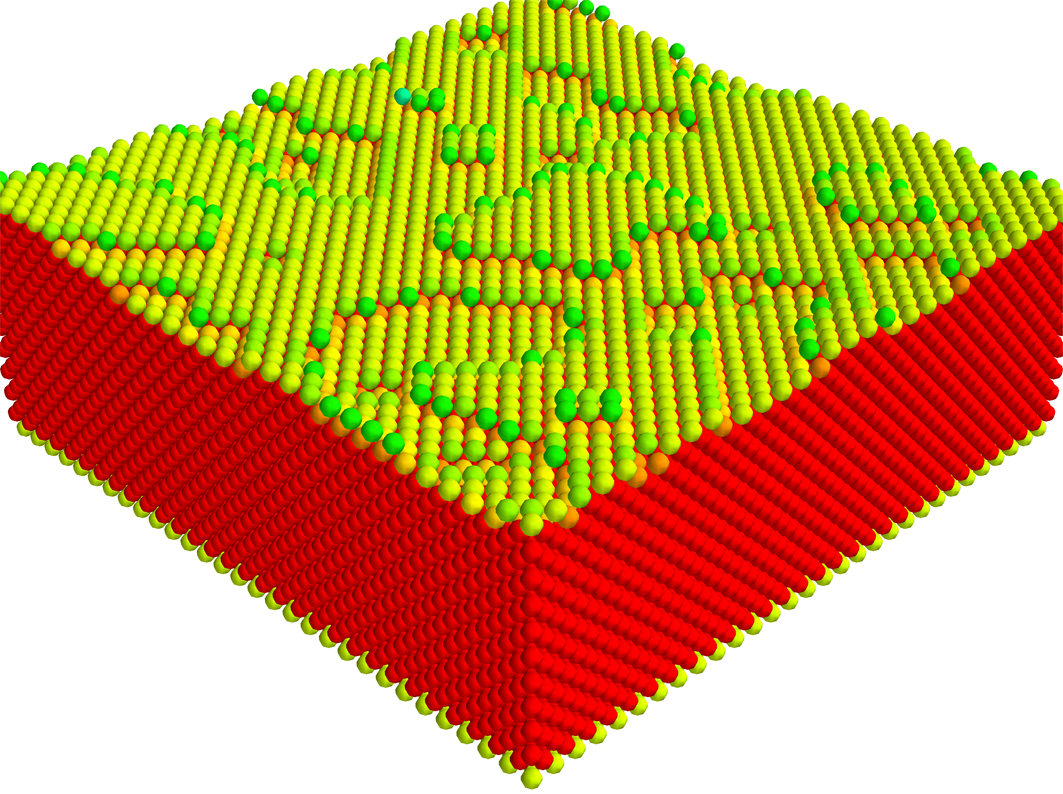}
                \caption{}
        \end{subfigure}

        \begin{subfigure}[b]{0.45\linewidth}
                \centering
                \includegraphics[width=\linewidth]{legend}
        \end{subfigure}
        \caption{Equilibrium morphology at t = 5 s of simulations with a) hopping as the only diffusion mechanism deposited at $-1000~\mathrm{A}~\mathrm{m}^{-2}$, b) hopping as the only diffusion mechanism deposited at $-100~\mathrm{A}~\mathrm{m}^{-2}$, c) all 3 diffusion mechanisms deposited at $-1000~\mathrm{A}~\mathrm{m}^{-2}$ and d) all 3 diffusion mechanisms deposited at $-100~\mathrm{A}~\mathrm{m}^{-2}$. Operating temperature is 300 K. Colors denote coordination number (blue to red in ascending order). The surface area of the substrates are $30a_{Cu} \times 30a_{Cu}$ ($\approx 120$ nm\textsuperscript{2}).\label{fig:diff_mech_comp}}
\end{figure*}

\subsection{Effect of Diffusion Mechanisms}\label{sec:effect_diff_mech}

In order to study the role of the surface diffusion mechanisms considered in KMC-EAM (Figure \ref{fig:diff_mech}) on deposit morphology, two sets of simulations were performed assuming that (i) hopping alone and (ii) all three modes operate.
Past KMC simulation studies typically include only the hopping mechanism \cite{Kaneko2006, Liu2009, Liu2013, Kaneko2013, Rubio2003}.
Restricting surface diffusion to only hopping precludes the possibility of adatoms diffusing from terraces in the deposit.
KMC-EAM simulations were carried out under these two conditions at current densities of $-100 ~\mathrm{A}~\mathrm{m}^{-2}$ and $-1000 ~\mathrm{A}~\mathrm{m}^{-2}$.
Equilibrium deposit configurations are shown in Figure \ref{fig:diff_mech_comp} and a distinct difference in deposit morphology is observed independent of current density.

Deposit morphologies predicted by KMC-EAM simulations with hopping-only show significant increase in roughness and cluster mean curvature.
Deposits simulated when all three diffusion mechanisms are included are less rough and distinct terraces are formed that are large compared to the previous case.
The average root-mean-squared roughness ($\overline{R}_{RMS}$) is calculated using \cite{Gadelmawla2002}:
\begin{equation}
R_{RMS} = \sqrt{\frac{1}{n}\sum_{i=1}^{n}\left(h_{i} - \overline{h}\right)^{2}},
\label{eq:roughness}
\end{equation}
where $h_{i}$ is the height of each surface atom, $\overline{h}$ is the average height, and $n$ is the number of surface atoms.
As shown in Table \ref{tab:roughness}, the roughness of the deposit when only hopping operates is significantly greater than when all surface diffusion modes are considered regardless of the current density.

\begin{table*}
\caption{Deposit cluster properties from Figure \ref{fig:diff_mech_comp} -- root-mean-squared roughness ($\overline{R}_{RMS}$), cluster area fraction ($\overline{A}$), average cluster perimeter ($\overline{P}$) and Euler characteristic ($\overline{\chi}$).}
\centering
\begin{tabular}{c c c c c c}
\firsthline
$i_{dep}$ & Diffusion & $\overline{R}_{RMS}$ & $\overline{A}$ & $\overline{P}$ & $\overline{\chi}$\\ 
(A m\textsuperscript{-2}) &Mechanisms & (nm) & & (nm) & (nm\textsuperscript{-1})\\
\hline
\multirow{2}{*}{$-1000$} & Hopping & 1.529 $\pm$ 0.004 & 0.14 $\pm$ 0.00 & 8.4 $\pm$ 0.3 & 1651.2 $\pm$ 17.4\\ & All & 0.872 $\pm$ 0.008 & 0.74 $\pm$ 0.00 & 101.6 $\pm$ 43.7 & 1620.8 $\pm$ 21.7\\
\hline
\multirow{2}{*}{$-100$} & Hopping & 1.580 $\pm$ 0.011 & 0.16 $\pm$ 0.00 & 9.8 $\pm$ 0.7 & 1607.7 $\pm$ 34.8\\ & All & 0.797 $\pm$ 0.022 & 0.80 $\pm$ 0.02 & 100.1 $\pm$ 29.3 & 1525.2 $\pm$ 265.1\\
\lasthline
\end{tabular}
\label{tab:roughness}
\end{table*}

In addition to surface roughness, the morphology of the deposit surface was quantified using the Minkowski measures \cite{Mantz2008}.
Three Minkowski measures are defined for a two-dimensional surface: surface area, perimeter and Euler characteristic.
The Euler characteristic is an integral measure of curvature over the cluster boundary.
To compute these morphological measures from a given deposit surface, they are converted to binary images using surface depth as image intensity.
Thus these morphological measures characterize the cluster morphology of the deposit.

Table \ref{tab:roughness} shows the morphological measures from the two sets of simulations.
The average cluster area fraction, $\overline{A}$, is the fraction of the total cluster surface area with respect to the total surface area.
The average cluster perimeter, $\overline{P}$, is the average perimeter of the clusters in the domain.
The average Euler characteristic, $\overline{\chi}$, is related to the total curvature of the cluster boundaries within the simulation domain.

At $-1000 ~\mathrm{A}~\mathrm{m}^{-2}$, the average cluster perimeter is lower when only hopping is involved than when three diffusion mechanisms are involved.
This corresponds to smaller clusters which is supported by a reduction in the average total cluster area.
Since step-edge atom exchange and atom exchange do not occur to level the surface and coalesce the clusters, this result is expected.
The measures obtained for deposition at $-100 ~\mathrm{A}~\mathrm{m}^{-2}$ are consistent with those obtained at the higher current. When three diffusion mechanisms are considered,  $\overline{A}$ is an order of magnitude greater than that obtained when only hopping is considered.

The average perimeters $\overline{P}$ for the two cases also agree with this trend by revealing smaller clusters when hopping is the only diffusion mechanism.
The Euler characteristic $\overline{\chi}$ and thus average curvature of the domains are similar, indicating that the curvature of the cluster boundaries is determined by minimization of the cluster/bulk interfacial energy and not specific diffusion mechanisms.

The deposit surface features support the qualitative observation made based on Figure \ref{fig:diff_mech_comp}.
When hopping is the only diffusion mechanism, the deposit has greater roughness and the individual clusters are smaller. 
The growth mode observed when three surface diffusion mechanisms are included is similar to that of Cu/Cu(100) homoepitaxial growth observed experimentally \cite{Ferron2000, Ernst1992}.

\subsection{Comparison of Equilibrium Deposits}\label{sec:validation_results}

The final set of simulations were performed over a range of initial conditions, current densities and temperatures using KMC-EAM.
Equilibrium deposit configurations from these KMC-EAM simulations were then used as initial conditions for MD-EAM simulations under commensurate conditions (temperature and ensemble).
Through relaxation of the approximations required to perform KMC, the MD-EAM simulations results were used to determine the validity of the KMC-EAM method for simulations of the electrodeposition process.
In all KMC-EAM simulations the occupancy fraction, $f_{s}=1.0$, corresponding to electrodeposition on atomically smooth copper crystal. 

\begin{figure}
\centering
\includegraphics[width=\linewidth]{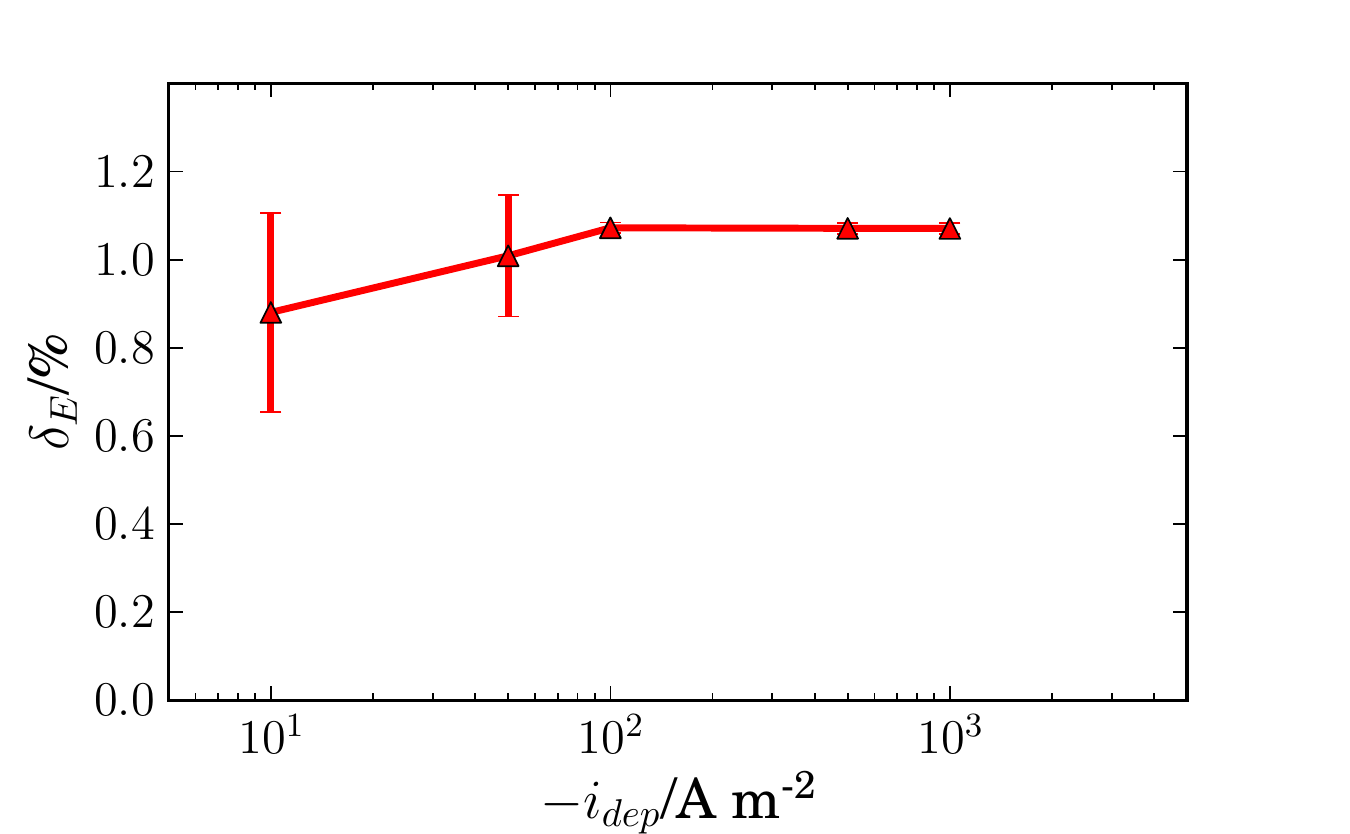}
\caption{Effect of current density on $\delta_{E}$ obtained by MD-EAM and KMC-EAM simulations at $300~\mathrm{K}$, $h_{s} = 1.1~\mathrm{nm}$ \ and $f_{s} = 1.0$.\label{fig:rate}}
\end{figure}

Figure \ref{fig:rate} shows the difference $\delta_{E}$ between the mean energy of the equilibrium deposit configurations from KMC-EAM and the potential energy component of the same relaxed configurations from MD-EAM.
The simulation results span current densities ranging from $-10$ to $-1000 ~\mathrm{A}~\mathrm{m}^{-2}$ at $300~\mathrm{K}$.
It is observed that $\delta_{E}$ is non-negligible but reasonable over the full range of applied current densities.
The trend of $\delta_{E}$ increasing with respect to current density is expected in that an increased deposition rate results in the formation of vacancies which result in lattice relaxations that are not accounted for in KMC-EAM.  
Furthermore, lattice relaxation at the deposit surface is also not accounted for, which contributes to $\delta_{E}$.

The difference in atom coordination number $\delta_{C}$ was also determined in order to compare the KMC-EAM equilibrium configurations to those of MD-EAM.
These plots are not shown since the values of $\delta_{C}$ were all negligible, less than $0.04\%$.
This implies that deposit morphology from KMC-EAM is almost identical to the average morphology from MD-EAM.
Furthermore, current density was not found to have a statistically significant effect on $\delta_{C}$.
Thus the difference in energy $\delta_{E}$ is primarily a consequence of the on-lattice approximation of KMC-EAM and not significant difference in the deposit morphology.

\begin{figure}
\centering
\includegraphics[width=\linewidth]{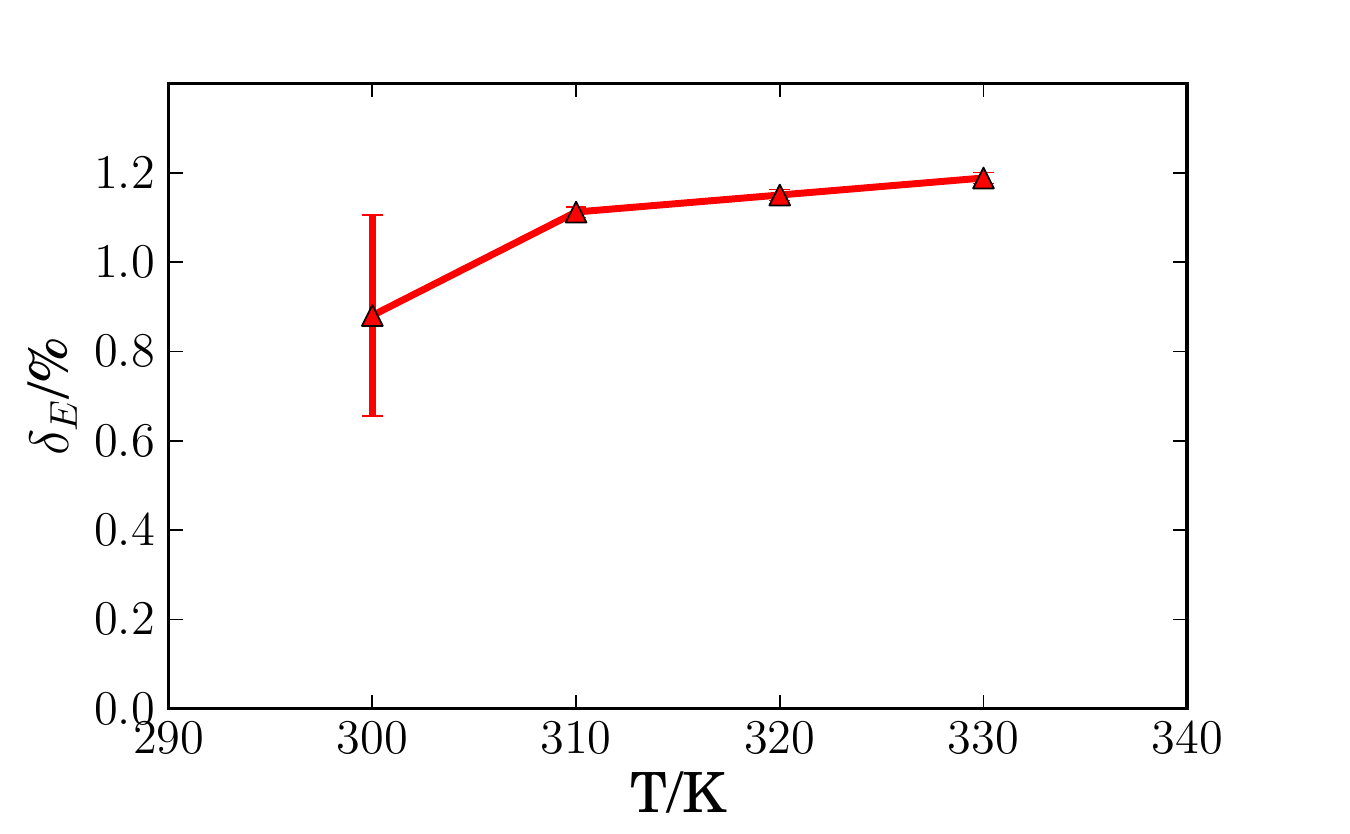}
\caption{Effect of temperature on $\delta_{E}$ obtained by MD-EAM and KMC-EAM simulations at $-10~\mathrm{A}~\mathrm{m}^{-2}$, $h_{s} = 1.1~\mathrm{nm}$ \ and $f_{s} = 1.0$.\label{fig:temp}}
\end{figure}

Figure \ref{fig:temp} shows the difference in energy $\delta_{E}$ between KMC-EAM and MD-EAM equilibrium deposit configurations for applied current density of $-10~\mathrm{A}~\mathrm{m}^{-2}$ over a range of temperatures $300-330~\mathrm{K}$.
A similar magnitude and trend of $\delta_{E}$ is observed as in the previous case with $\delta_{E}$ being non-negligible but reasonable over the full range of operating temperatures.
The values of $\delta_{C}$ are again negligible and thus not shown.
The results can be interpreted in the same way as before, but now increasing temperature results in the increased formation of vacancies and also increased lattice strain in the MD simulations.
The trend is slightly steeper than what results from the increase of current density, which implies that the KMC-EAM method will monotonically decrease in accuracy as temperature is increased.
The ranges of operating temperatures used in this work are typical for electrodeposition processes, thus within this range the KMC-EAM performs adequately with respect to comparison with MD-EAM.

The final metric used to evaluate the deposit configuration predicted by KMC-EAM is the RMS displacement of atoms from their starting positions obtained from the metastable configuration of KMC-EAM to reach their final positions as computed by MD-EAM.
The RMS displacement value for equilibrium single crystal copper deposits is reported to be $0.0113~\mathrm{nm}$ at $300~ \mathrm{K}$ \cite{Yang1991}.
The range of RMS displacement values for the KMC-EAM simulations with current density varied at 300 K was found to range between $0.019-0.021~\mathrm{nm}$.
For the set of simulations in which temperature was varied, the RMS displacement values ranged between $0.019-0.023~\mathrm{nm}$.
These results indicate that the equilibrium configuration predicted by KMC-EAM simulations essentially equivalent to that of MD-EAM.
Furthermore, the RMS displacement values appear to be only slightly affected by the operating conditions, which supports the interpretation of $\delta_{E}$ and $\delta_{C}$ trends discussed previously.

\section{Conclusions}\label{sec:concl}

A kinetic Monte Carlo methodology which uses the embedded-atom method potential and includes collective diffusion mechanisms (KMC-EAM) has been developed.
This methodology was applied to the simulation of galvanostatic electrodeposition of metals onto a single-crystal substrate of the same species.  
The average energy per atom and coordination number of equilibrium configurations from KMC-EAM were validated using MD simulation. 
The KMC-EAM was found to be accurate for deposition current density and temperature values relevant to experimental conditions.
Furthermore, the KMC-EAM accurately describes the nanoscale structure of the metal deposit through direct representation of the constituent atoms, unles the SOS and SBS methods.

In addition to analysis of equilibrium configurations, the effects of surface diffusion mechanisms (hopping, atom exchange and step-edge exchange) and diffusion kinetics were also studied.
Results show that the inclusion of collective diffusion mechanisms (atom exchange and step-edge exchange), in addition to nearest-neighbor hopping, 
were required to predict deposit configurations in agreement with both MD-EAM simulations and experimental results for Cu/Cu(100) homoepitaxy.
The inclusion of the three surface diffusion mechanisms resulted in quantitatively smoother deposits, as reflected by surface morphology measures  -- roughness, cluster perimeter and cluster area.

The diffusion kinetics observed indicated that the step-edge exchange mechanism was active predominantly during the deposition process, while hopping and atom-exchange continued following the cessation of electrodeposition.
In summary, the presented KMC-EAM method is shown to provide an accurate representation of the electrodeposition process which is able to perform simulations on experimentally relevant length (microns) and time (seconds).

\section*{Acknowledgments}
  This research was supported by the Natural Sciences and Engineering Research Council (NSERC) of Canada and the facilities of the Shared Hierarchical Academic Research Computing Network (SHARCNET:www.sharcnet.ca).
  The authors also thank Robert Suderman for his assistance with analysis of images of the deposit morphologies.

\bibliographystyle{elsarticle-num}
\bibliography{deposition,computational}

\begin{thebibliography}{10}
\expandafter\ifx\csname url\endcsname\relax
  \def\url#1{\texttt{#1}}\fi
\expandafter\ifx\csname urlprefix\endcsname\relax\def\urlprefix{URL }\fi
\expandafter\ifx\csname href\endcsname\relax
  \def\href#1#2{#2} \def\path#1{#1}\fi

\bibitem{Andricacos1998}
P.~C. Andricacos, C.~Uzoh, J.~O. Dukovic, J.~Horkans, H.~Deligianni, Damascene
  copper electroplating for chip interconnections, IBM J. Res. Develop. 42
  (1998) 567 --574.
\newblock \href {http://dx.doi.org/10.1147/rd.425.0567}
  {\path{doi:10.1147/rd.425.0567}}.

\bibitem{Durkan2000}
C.~Durkan, M.~E. Welland, Size effects in the electrical resistivity of
  polycrystalline nanowires, Phys. Rev. B 61 (2000) 14215--14218.
\newblock \href {http://dx.doi.org/10.1103/PhysRevB.61.14215}
  {\path{doi:10.1103/PhysRevB.61.14215}}.

\bibitem{Hau-Riege2001}
C.~S. Hau-Riege, C.~V. Thompson, Electromigration in {C}u interconnects with
  very different grain structures, Appl. Phys. Lett. 78 (2001) 3451--3453.
\newblock \href {http://dx.doi.org/10.1063/1.1355304}
  {\path{doi:10.1063/1.1355304}}.

\bibitem{Daw1984}
M.~S. Daw, M.~I. Baskes, Embedded-atom method: {D}erivation and application to
  impurities, surfaces, and other defects in metals, Phys. Rev. B 29 (1984)
  6443--6453.
\newblock \href {http://dx.doi.org/10.1103/PhysRevB.29.6443}
  {\path{doi:10.1103/PhysRevB.29.6443}}.

\bibitem{Daw1993}
M.~S. Daw, S.~M. Foiles, M.~I. Baskes, The embedded-atom method: a review of
  theory and applications, Mater. Sci. Rep. 9 (1993) 251--310.
\newblock \href {http://dx.doi.org/10.1016/0920-2307(93)90001-U}
  {\path{doi:10.1016/0920-2307(93)90001-U}}.

\bibitem{Adams1989}
J.~B. Adams, S.~M. Foiles, W.~G. Wolfer, Self-diffusion and impurity diffusion
  of fcc metals using the five-frequency model and the {E}mbedded {A}tom
  {M}ethod, J. Mater. Res. 4 (1989) 102--112.
\newblock \href {http://dx.doi.org/10.1557/JMR.1989.0102}
  {\path{doi:10.1557/JMR.1989.0102}}.

\bibitem{Foiles1987}
S.~M. Foiles, M.~I. Baskes, C.~F. Melius, M.~S. Daw, Calculation of hydrogen
  dissociation pathways on nickel using the embedded atom method, J.
  Less-Common Met. 130 (1987) 465--473.
\newblock \href {http://dx.doi.org/10.1016/0022-5088(87)90144-5}
  {\path{doi:10.1016/0022-5088(87)90144-5}}.

\bibitem{Antczak2010}
G.~Antczak, G.~Ehrlich, Surface Diffusion, Cambridge University Press, New York
  City, 2010.

\bibitem{Mariscal2007}
M.~Mariscal, E.~Leiva, K.~P\"{o}tting, W.~Schmickler, The structure of
  electrodeposits - a computer simulation study, Appl. Phys. A 87 (2007)
  385--389.
\newblock \href {http://dx.doi.org/10.1007/s00339-007-3915-y}
  {\path{doi:10.1007/s00339-007-3915-y}}.

\bibitem{Voter1997}
A.~F. Voter, Hyperdynamics: Accelerated molecular dynamics of infrequent
  events, Phys. Rev. Lett. 78 (1997) 3908--3911.
\newblock \href {http://dx.doi.org/10.1103/PhysRevLett.78.3908}
  {\path{doi:10.1103/PhysRevLett.78.3908}}.

\bibitem{Voter1997-2}
A.~F. Voter, A method for accelerating the molecular dynamics simulation of
  infrequent events, J. Chem. Phys. 106~(11) (1997) 4665--4677.
\newblock \href {http://dx.doi.org/10.1063/1.473503}
  {\path{doi:10.1063/1.473503}}.

\bibitem{Sorensen2000-TAD}
M.~R. Sorensen, A.~F. Voter, Temperature-accelerated dynamics for simulation of
  infrequent events, J. Chem. Phys. 112 (2000) 9599--9606.
\newblock \href {http://dx.doi.org/10.1063/1.481576}
  {\path{doi:10.1063/1.481576}}.

\bibitem{Voter2002}
A.~F. Voter, F.~Montalenti, T.~C. Germann, Extending the time scale in
  atomistic simulaton of materials, Annu. Rev. Mater. Res. 32~(1) (2002)
  321--346.
\newblock \href
  {http://arxiv.org/abs/http://www.annualreviews.org/doi/pdf/10.1146/annurev.matsci.32.112601.141541}
  {\path{arXiv:http://www.annualreviews.org/doi/pdf/10.1146/annurev.matsci.32.112601.141541}},
  \href {http://dx.doi.org/10.1146/annurev.matsci.32.112601.141541}
  {\path{doi:10.1146/annurev.matsci.32.112601.141541}}.

\bibitem{Fichthorn1991}
K.~A. Fichthorn, W.~H. Weinberg, Theoretical foundations of dynamical monte
  carlo simulations, J. Chem. Phys. 95 (1991) 1090--1096.
\newblock \href {http://dx.doi.org/10.1063/1.461138}
  {\path{doi:10.1063/1.461138}}.

\bibitem{Gilmer1972}
G.~H. Gilmer, P.~Bennema, Simulation of crystal growth with surface diffusion,
  J. Appl. Phys. 43 (1972) 1347--1360.
\newblock \href {http://dx.doi.org/10.1063/1.1661325}
  {\path{doi:10.1063/1.1661325}}.

\bibitem{Zheng2008}
Z.~Zheng, R.~M. Stephens, R.~D. Braatz, R.~C. Alkire, L.~R. Petzold, A hybrid
  multiscale kinetic {M}onte {C}arlo method for simulation of copper
  electrodeposition, J. Comput. Phys. 227 (2008) 5184--5199.
\newblock \href {http://dx.doi.org/10.1016/j.jcp.2008.01.056}
  {\path{doi:10.1016/j.jcp.2008.01.056}}.

\bibitem{Rusli2007}
E.~Rusli, F.~Xue, T.~O. Drews, P.~M. Vereecken, P.~Andricacos, H.~Deligianni,
  R.~D. Braatz, R.~C. Alkire, Effect of additives on shape evolution during
  electrodeposition {II}. {P}arameter estimation from roughness evolution
  experiments, J. Electrochem. Soc. 154 (2007) D584--D597.
\newblock \href {http://dx.doi.org/10.1149/1.2772425}
  {\path{doi:10.1149/1.2772425}}.

\bibitem{Rubio2003}
J.~E. Rubio, M.~Jaraiz, I.~Martin-Bragado, J.~M. Hernandez-Mangas, J.~Barbolla,
  G.~H. Gilmer, Atomistic monte carlo simulations of three-dimensional
  polycrystalline thin films, J. Appl. Phys. 94 (2003) 163--168.
\newblock \href {http://dx.doi.org/10.1063/1.1577814}
  {\path{doi:10.1063/1.1577814}}.

\bibitem{Stephens2007}
R.~M. Stephens, R.~C. Alkire, Simulation of kinetically limited nucleation and
  growth at monatomic step edges, J. Electrochem. Soc. 154 (2007) D418--D426.
\newblock \href {http://dx.doi.org/10.1149/1.2746569}
  {\path{doi:10.1149/1.2746569}}.

\bibitem{Wang2000}
Z.~Wang, Y.~Li, J.~B. Adams, Kinetic lattice {M}onte {C}arlo simulation of
  facet growth rate, Surf. Sci. 450 (2000) 51--63.
\newblock \href {http://dx.doi.org/10.1016/S0039-6028(99)01250-9}
  {\path{doi:10.1016/S0039-6028(99)01250-9}}.

\bibitem{Liu2009}
J.~Liu, C.~Liu, P.~P. Conway, Kinetic {M}onte {C}arlo simulation of kinetically
  limited copper electrocrystallization on an atomically even surface,
  Electrochim. Acta 54 (2009) 6941--6948.
\newblock \href {http://dx.doi.org/10.1016/j.electacta.2009.07.019}
  {\path{doi:10.1016/j.electacta.2009.07.019}}.

\bibitem{Liu2009EC}
J.~Liu, C.~Liu, P.~P. Conway, Kinetic {M}onte {C}arlo simulation of
  electrodeposition of polycrystalline {C}u, Electrochem. Commun. 11 (2009)
  2207--2211.
\newblock \href {http://dx.doi.org/10.1016/j.elecom.2009.09.032}
  {\path{doi:10.1016/j.elecom.2009.09.032}}.

\bibitem{Liu2013}
J.~Liu, C.~Liu, P.~P. Conway, Kinetic {M}onte {C}arlo simulation of the
  electrodeposition of polycrystalline copper: Effects of substrates and
  deposition parameters on the microstructure of deposits, Electrochim. Acta 97
  (2013) 132 -- 142.
\newblock \href {http://dx.doi.org/10.1016/j.electacta.2013.02.112}
  {\path{doi:10.1016/j.electacta.2013.02.112}}.

\bibitem{Kaneko2006}
Y.~Kaneko, Y.~Hiwatari, K.~Ohara, F.~Asa, Monte {C}arlo simulation of damascene
  electroplating: effects of additives, Mol. Simul. 32 (2006) 1227--1232.
\newblock \href {http://dx.doi.org/10.1080/08927020601067540}
  {\path{doi:10.1080/08927020601067540}}.

\bibitem{Kaneko2009}
Y.~Kaneko, S.~Nishimura, Y.~Hiwatari, K.~Ohara, F.~Asa, Monte {C}arlo and
  molecular dynamics studies of the effects of additives in electrodeposition,
  J. Korean Phys. Soc. 54 (2009) 1207--1211.
\newblock \href {http://dx.doi.org/10.3938/jkps.54.1207}
  {\path{doi:10.3938/jkps.54.1207}}.

\bibitem{Kaneko2013}
Y.~Kaneko, Y.~Hiwatari, K.~Ohara, F.~Asa, Kinetic {M}onte {C}arlo simulation of
  three-dimensional shape evolution with void formation using solid-by-solid
  model: Application to via and trench filling, Electrochim. Acta 100 (2013)
  321--328.
\newblock \href {http://dx.doi.org/10.1016/j.electacta.2013.01.076}
  {\path{doi:10.1016/j.electacta.2013.01.076}}.

\bibitem{Gilmer2000}
G.~H. Gilmer, H.~Huang, T.~D. de~la Rubia, J.~D. Torre, F.~Baumann, Lattice
  monte carlo models of thin film deposition, Thin Solid Films 365 (2000) 189
  -- 200.
\newblock \href {http://dx.doi.org/10.1016/S0040-6090(99)01057-3}
  {\path{doi:10.1016/S0040-6090(99)01057-3}}.

\bibitem{Gimenez2002}
M.~C. Gimenez, M.~G. {Del Popolo}, E.~P.~M. Leiva, Kinetic {M}onte {C}arlo
  study of electrochemical growth in a heteroepitaxial system, Langmuir 18
  (2002) 9087--9094.
\newblock \href {http://dx.doi.org/10.1021/la020505y}
  {\path{doi:10.1021/la020505y}}.

\bibitem{Gimenez2003}
M.~C. Gimenez, M.~Cecilia, E.~P.~M. Leiva, Comparative {M}onte {C}arlo study of
  monolayer growth in a heteroepitaxial system in the presence of surface
  defects, Langmuir 19 (2003) 10538--10549.
\newblock \href {http://dx.doi.org/10.1021/la035350o}
  {\path{doi:10.1021/la035350o}}.

\bibitem{Oviedo2005}
O.~A. Oviedo, M.~I. Rojas, E.~P.~M. Leiva, Off lattice {M}onte-{C}arlo
  simulations of low-dimensional surface defects and metal deposits on
  {P}t(111), Electrochem. Commun. 7 (2005) 472--476.
\newblock \href {http://dx.doi.org/10.1016/j.elecom.2005.03.002}
  {\path{doi:10.1016/j.elecom.2005.03.002}}.

\bibitem{Huang2009}
Y.~Y. Huang, Y.~C. Zhou, Y.~Pan, Simulation of kinetically limited growth of
  electrodeposited polycrystalline {N}i films, Physica E 41 (2009) 1673--1678.
\newblock \href {http://dx.doi.org/10.1016/j.physe.2009.06.001}
  {\path{doi:10.1016/j.physe.2009.06.001}}.

\bibitem{Chatterjee2007}
A.~Chatterjee, D.~G. Vlachos, An overview of spatial microscopic and
  accelerated kinetic {M}onte {C}arlo methods, J. Comput. Aided Mater. Des. 14
  (2007) 253--308.
\newblock \href {http://dx.doi.org/10.1007/s10820-006-9042-9}
  {\path{doi:10.1007/s10820-006-9042-9}}.

\bibitem{Gardiner1985}
C.~W. Gardiner, Handbook of Stochastic Methods, 2nd Edition, Springer-Verlag,
  1985.

\bibitem{Gillespie2007}
D.~T. Gillespie, Stochastic simulation of chemical kinetics, Annu. Rev. Phys.
  Chem. 58 (2007) 35--55.
\newblock \href {http://dx.doi.org/10.1146/annurev.physchem.58.032806.104637}
  {\path{doi:10.1146/annurev.physchem.58.032806.104637}}.

\bibitem{Mattsson1959}
E.~Mattsson, J.~O. Bockris, Galvanostatic studies of the kinetics of deposition
  and dissolution in the copper + copper sulphate system, Trans. Faraday Soc.
  55 (1959) 1586--1601.
\newblock \href {http://dx.doi.org/10.1039/TF9595501586}
  {\path{doi:10.1039/TF9595501586}}.

\bibitem{Conway1961}
B.~E. Conway, J.~O. Bockris, On the calculation of potential energy profile
  diagrams for processes in electrolytic metal deposition, Electrochim. Acta 3
  (1961) 340 -- 366.
\newblock \href {http://dx.doi.org/10.1016/0013-4686(61)85009-3}
  {\path{doi:10.1016/0013-4686(61)85009-3}}.

\bibitem{Wolf1992}
D.~Wolf, Atomic-level geometry of crystalline interfaces, in: D.~Wolf, S.~Yip
  (Eds.), Materials interfaces: {A}tomic-level structure and properties, 1st
  Edition, Chapman and Hall, 1992, Ch.~1, pp. 1--57.

\bibitem{Budevski1996}
E.~Budevski, G.~Staikov, W.~J. Lorenz, Electrochemical phase formation and
  growth: An Introduction to the Initial Stages of Metal Deposition, Advances
  in Electrochemical Science and Engineering, VCH, Weinheim, Germany, 1996.

\bibitem{Battaile2008}
C.~C. Battaile, The kinetic {M}onte {C}arlo method: Foundation, implementation,
  and applications, Comput. Methods Appl. Mech. Engrg. 197~(41-42) (2008) 3386
  -- 3398.
\newblock \href {http://dx.doi.org/10.1016/j.cma.2008.03.010}
  {\path{doi:10.1016/j.cma.2008.03.010}}.

\bibitem{Plimpton2009}
S.~Plimpton, C.~Battaile, M.~Chandross, L.~Holm, A.~Thompson, V.~Tikare,
  G.~Wagner, E.~Webb, X.~Zhou, C.~G. Cardona, A.~Slepoy,
  \href{http://spparks.sandia.gov/pdf/sand09.pdf}{Crossing the mesoscale
  no-man's land via parallel kinetic {M}onte {C}arlo}, Sandia report
  {SAND}2009-6226, Sandia National Laboratories (2009).
\newline\urlprefix\url{http://spparks.sandia.gov/pdf/sand09.pdf}

\bibitem{Gibson2000}
M.~A. Gibson, J.~Bruck, Efficient exact stochastic simulation of chemical
  systems with many species and many channels, J. Phys. Chem. A 104~(9) (2000)
  1876--1889.
\newblock \href {http://dx.doi.org/10.1021/jp993732q}
  {\path{doi:10.1021/jp993732q}}.

\bibitem{Plimpton1995}
S.~Plimpton, Fast parallel algorithms for short-range molecular dynamics, J.
  Comput. Phys. 117~(1) (1995) 1 -- 19.
\newblock \href {http://dx.doi.org/10.1006/jcph.1995.1039}
  {\path{doi:10.1006/jcph.1995.1039}}.

\bibitem{Vazquez2013}
J.~Vazquez-Arenas, M.~Pritzker, M.~Fowler, Kinetic and hydrodynamic
  implications of 1-{D} and 2-{D} models for copper electrodeposition under
  mixed kinetic-mass transfer control, Electrochim. Acta 89 (2013) 717 -- 725.
\newblock \href {http://dx.doi.org/10.1016/j.electacta.2012.11.024}
  {\path{doi:10.1016/j.electacta.2012.11.024}}.

\bibitem{Bard2001}
A.~J. Bard, L.~R. Faulkner, Electrochemical Methods: Fundamentals and
  Applications, 2nd Edition, John Wiley and Sons, New York, New York, 2001.

\bibitem{Gadelmawla2002}
E.~Gadelmawla, M.~Koura, T.~Maksoud, I.~Elewa, H.~Soliman, Roughness
  parameters, J. Mater. Process. Technol. 123~(1) (2002) 133 -- 145.
\newblock \href {http://dx.doi.org/10.1016/S0924-0136(02)00060-2}
  {\path{doi:10.1016/S0924-0136(02)00060-2}}.

\bibitem{Mantz2008}
H.~Mantz, K.~Jacobs, K.~Mecke, Utilizing {M}inkowski functionals for image
  analysis: a marching square algorithm, J. Stat. Mech: Theory Exp. 2008~(12)
  (2008) P12015.
\newblock \href {http://dx.doi.org/10.1088/1742-5468/2008/12/P12015}
  {\path{doi:10.1088/1742-5468/2008/12/P12015}}.

\bibitem{Ferron2000}
J.~Ferr\'{o}n, L.~G\'{o}mez, J.~Gallego, J.~Camarero, J.~E. Prieto, V.~Cros,
  A.~L. V\'{a}zquez~de Parga, J.~J. de~Miguel, R.~Miranda, Influence of
  surfactants on atomic diffusion, Surf. Sci. 459 (2000) 135 -- 148.
\newblock \href {http://dx.doi.org/10.1016/S0039-6028(00)00459-3}
  {\path{doi:10.1016/S0039-6028(00)00459-3}}.

\bibitem{Ernst1992}
H.-J. Ernst, F.~Fabre, J.~Lapujoulade, Growth of {C}u on {C}u(100), Surf. Sci.
  275 (1992) L682 -- L684.
\newblock \href {http://dx.doi.org/10.1016/0039-6028(92)90641-I}
  {\path{doi:10.1016/0039-6028(92)90641-I}}.

\bibitem{Yang1991}
L.~Yang, T.~S. Rahman, M.~S. Daw, Surface vibrations of {A}g(100) and
  {C}u(100): A molecular-dynamics study, Phys. Rev. B 44 (1991) 13725--13733.
\newblock \href {http://dx.doi.org/10.1103/PhysRevB.44.13725}
  {\path{doi:10.1103/PhysRevB.44.13725}}.

\end{thebibliography}

\end{document}